\documentclass[a4paper,11pt]{article}
\usepackage{jinstpub} 
\usepackage{lineno}
\usepackage{subfig}
\usepackage[export]{adjustbox}
\usepackage{wasysym}

\usepackage{booktabs}
\usepackage{siunitx}
\usepackage{xspace}

\newcommand{\LNtwo}{LN$_2$\xspace}
\newcommand{\LHe}{LHe\xspace}
\newcommand{\LAr}{LAr\xspace}

\newcommand{\Eb}{E_{\mathrm{b}}}
\newcommand{\Vb}{V_{\mathrm{b}}}

\title{Comparative measurements of DC electrical breakdown distributions in liquid nitrogen, liquid helium, and liquid argon}

\author[a,1]{N.~S. Phan,\note{Corresponding author.}}
\author[a]{S.~M.~Clayton,}
\author[b]{R.~Gautam,}
\author[a]{T.~M.~Ito,}
\author[b]{L.~Kadlec,}
\author[a]{C.~M.~O'Shaughnessy,}
\author[a]{T.~J.~Schaub}

\affiliation[a]{Los Alamos National Laboratory, Los Alamos, New Mexico 87545, U.S.A.}

\affiliation[b]{Department of Physics and Astronomy, Valparaiso University, Valparaiso, Indiana 46383, U.S.A.}

\emailAdd{nphan@protonmail.com}

\abstract{
We present a comparative study of DC electrical breakdown in liquid nitrogen, liquid helium, and liquid argon using a common cryostat, electrode assembly, voltage-ramp protocol, and analysis procedure. This common experimental configuration minimizes systematic differences that often complicate comparisons among cryogenic liquids, including variations in electrode geometry, stressed area, surface finish, and gap spacing. For each liquid, breakdown was characterized as a statistical distribution rather than by a single characteristic voltage, allowing the stochastic nature of the process to be examined. Measurements were performed both near saturated vapor pressure and under modest pressurization to assess the influence of bubble formation and related thermodynamic effects. Liquid argon exhibited a non-stationary ``turn-on'' behavior, with breakdown voltage increasing over repeated discharges, consistent with evolution of the local impurity environment or electrode surface state. Near saturated vapor pressure, liquid helium showed a bimodal breakdown distribution, with a lower-field component consistent with bubble-mediated processes. Liquid nitrogen exhibited higher breakdown fields than liquid argon and fields comparable to those measured in pressurized liquid helium, within the observed run-to-run variation associated with surface state. Taken together, these results support a comparative framework in which
breakdown strength reflects both thermodynamic stability against bubble formation and the transport properties of the dominant negative charge
carrier.
}


\begin{document}
\maketitle
\flushbottom


\section{Introduction}\label{sec:intro}
 
Modern technological and fundamental research applications are increasingly characterized by the integration of cryogenics and high-voltage (HV) engineering. In disciplines ranging from fundamental physics~\cite{Rebel2014} to HV power transmission~\cite{Minnich1969}, liquefied gases—most notably liquid nitrogen (LN$_2$), liquid helium (LHe), liquid argon (LAr), and liquid xenon (LXe)—serve a critical, multifunctional role.  These fluids function simultaneously as essential coolants, required to establish and maintain superconductivity, and as the primary dielectric insulating medium.  This capability is indispensable for the advancement of high-temperature superconducting (HTS) power cables~\cite{Masuda2005, Yazdani-Asrami2022, Paramane2023}, superconducting fault current limiters (SFCLs)~\cite{Gray1978, Noe2007, Alam2018, Guilherme2022}, and superconducting magnetic energy storage (SMES) systems~\cite{Boom1972, Hassenzahl1983, Luongo1996, Adetokun2022}. Furthermore, cryogenic systems are integral to modern and proposed particle accelerators, where liquid-helium refrigeration enables the operation of superconducting radio-frequency cavities and magnets ~\cite{Padamsee2001,ILCTDR}. Large-scale noble-liquid time projection chambers, by contrast, use the cryogenic liquid directly as the active detector and dielectric medium and require stable electric fields for charge drift, extraction, and collection.
Such capabilities are fundamental to experiments investigating dark matter~\cite{ArDM2010, Darkside2011, LZ2015, PandaX2018, XENON2024}, neutrino physics~\cite{Acciarri2017, Abi2020, nEXO2018, EX0200}, and fundamental symmetries via the measurement of electric dipole moments (EDMs)~\cite{Baker2010, Ahmed2019, Ito2025}.

However, operating these systems presents challenges because high electric fields can lead to stochastic breakdown. Despite existing theoretical and experimental studies on HV breakdown in cryogenic media, a comprehensive and cohesive understanding remains elusive. This can be attributed to the complex physics of breakdown, which is dependent on a multitude of seemingly disparate parameters. These parameters include electrode material, size, shape, and surface condition, as well as gap distance, liquid type, purity, temperature, pressure, voltage waveform and polarity, voltage ramp rate, and impulse time. As such, the complexity arising from the interactions of these factors can obscure the underlying physical mechanisms.

To simplify the problem, a cross-comparison of the breakdown behavior in different liquids can be performed to identify common underlying principles. This comparison is particularly valuable when key experimental parameters are
controlled to reduce systematic effects. For example, variations in electrode size, shape, material composition, and surface polish are common between different experiments, preventing straightforward comparisons.

Nonetheless, a controlled characterization of the relative breakdown behavior
of these liquids yields practical benefits. In research and development,
lower-cost cryogens such as \LNtwo can sometimes serve as useful surrogate
media for commissioning HV designs before operation in more demanding
cryogens such as \LHe. Such substitution is not exact, but it can reduce cost,
time, and operational complexity when the relevant limitations are understood.

A complementary strategy is to acquire sufficiently large samples to characterize the full breakdown-voltage distributions rather than rely on sparse measurements. This improves sensitivity to distributional features that may be obscured in small datasets and permits direct examination of the
lower-probability tails. Such tails can be more relevant to operational risk and stability than the mean behavior alone.

Systematic differences in experimental design and operating procedures in previous studies, together with the limited statistical samples often reported,
have contributed to apparently conflicting results. To reduce these ambiguities, this work investigates DC HV breakdown in LHe, LAr, and LN$_2$ using a common cryostat and electrode assembly. For each condition, we acquired a substantial sample of breakdown events, enabling a more direct
comparison of the resulting distributions. The principal contribution of this work is therefore not a single absolute breakdown-strength value for each cryogen, but a controlled comparison of the full breakdown distributions obtained with the same electrode pair and experimental procedure. This distinction is important because the breakdown probability relevant for detector or power-system design is often determined by the lower tail of the distribution rather than by the mean alone. Throughout the analysis, we report both central values and distributional features, and we distinguish measured trends from physical interpretations that remain model-dependent. 

The paper is structured as follows: Section~\ref{sec:review} provides a brief review of some key parameters that influence electrical breakdown in cryogenic liquids. The complications associated with controlling these parameters make direct comparisons of results from different experiments challenging. Section~\ref{sec:setup} outlines the experimental apparatus and measurement procedure, while Section~\ref{sec:results} presents the breakdown data collected for each cryogen. Section~\ref{sec:discussion} presents a comparative analysis, emphasizing the empirical
relationship between negative-carrier transport and breakdown strength and the role of bubble formation in LHe, which may help reconcile discrepancies
in previous studies. The paper concludes with a summary of the results in Section~\ref{sec:summary}.


\section{Factors affecting dielectric breakdown}\label{sec:review}

Electrical breakdown in cryogenic liquids is governed by a set of coupled transport, thermodynamic, and interfacial processes. Although the applied electric field is the primary driver, the measured breakdown distribution depends on the dominant negative charge carrier, the impurity content, the liquid thermodynamic state, the electrode surface, and electrode geometry. These dependencies complicate comparisons between independent experiments, in which the stressed area, gap spacing, surface finish, pressure, purity, and measurement procedure are often not equivalent. The following summary identifies the principal parameters relevant to the present comparison and motivates the use of a common electrode assembly, cryostat, voltage-ramp procedure, and statistical treatment.


\subsection{Charge transport and impurities} \label{sec:charge-transport-impurities}

The transport properties of the dominant negative charge carrier influence the rate at which the injected charge gains energy from the field and participates in avalanche, streamer, thermal, or surface-assisted precursor processes. In LAr, the negative carrier is a quasi-free electron with high mobility~\cite{Miller1968}, whereas in LN$_2$ the carrier is much less mobile and is effectively localized as a negative ion~\cite{Gee1985}. In LHe, electrons form localized vacuum cavities, also known as "electron
bubbles," with mobilities intermediate between these two cases~\cite{Meyer1962}. This classification is necessarily approximate, since the relevant mobility at breakdown may differ from the low-field transport value and may depend on the pressure, electric field, and impurity concentration. Nevertheless, previous measurements indicate a useful empirical trend: liquids with high negative-carrier mobility often exhibit lower breakdown fields than liquids in which the carrier is localized or strongly scattered~\cite{Swan1960,Mathes1967,Tvrznikova2019}.

Impurities can modify this behavior by altering both charge transport and local field conditions. Electronegative contaminants capture quasi-free electrons and form less mobile negative ions, thereby reducing the effective mobility of the negative charge carrier. In LAr, oxygen contamination has been shown to substantially increase the breakdown strength~\cite{Swan1961}. In cryogenic liquids, contaminants may also freeze onto electrode surfaces or remain localized near the high-field region after discharge events, coupling bulk purity to surface state and discharge history. For this reason, purity should be regarded as a local experimental condition rather than solely as a property of the source liquid or gas.

\subsection{Thermodynamic state and vapor bubble formation} \label{sec:thermal-bubble}

The thermodynamic stability of the liquid is central to the breakdown behavior near saturated vapor pressure. Vapor bubbles have a lower dielectric strength than the surrounding liquid and can provide favorable sites for charge multiplication, streamer propagation, or field-emission-assisted growth. Local vapor formation can be driven by residual heat load, field emission from cathode asperities, or other surface-mediated heating processes. Subcooling or pressurization suppresses bubble nucleation and growth and therefore generally increases dielectric stability when bubble-mediated processes are important~\cite{Na2010, Blaz2011, Nishimachi2012, Hayakawa2014}.

This effect is expected to be especially pronounced in LHe because of its low latent heat of vaporization. A heat input that produces little vapor in LN$_2$ or LAr can generate a much larger fractional vapor response in LHe, particularly near saturated vapor pressure~\cite{Donnelly1998,Jacobsen1986,Chen1975}. Consequently, LHe breakdown measurements are especially sensitive to pressure, heat load, and surface-assisted nucleation. Pressure-dependent breakdown has been reported in LHe~\cite{Mathes1967, Burnier1976, Yoshino1982, Phan2021}, liquid hydrogen (LH$_2$)~\cite{Jefferies1970, Burnier1976}, and LN$_2$~\cite{Jefferies1970, Kawashima1974, Delfino1979, Shiraishi1981, Yoshino1982, Nishimachi2012, Blaz2012}, while LAr~\cite{Yoshino1982, Tvrznikova2019} and LXe~\cite{Tvrznikova2019} generally show a more moderate pressure response under comparable HV conditions. The pressure dependence observed in the present measurements provides a gauge of the extent to which bubble formation contributes to the breakdown mechanism.

\subsection{Electrode surface, geometry, and stressed volume} \label{sec:electrode-surface}
Electrode properties determine the local field distribution and the probability that breakdown is initiated at a weak site. Surface asperities, microprotrusions, adsorbed contaminants, oxides, and frozen particulates can enhance the local electric field, promote electron injection, or lower the barrier for heterogeneous bubble nucleation~\cite{May1981,Shin2012}. Electrical conditioning may remove or modify such sites, but it can also change the surface in ways that are difficult to reproduce. Consequently, electrode history can contribute to run-to-run variation even when the macroscopic geometry is unchanged.

Geometry and scale effects are similarly important. Sharp edges and small radii of curvature concentrate the field, while larger stressed areas increase the probability that an initiating defect or impurity site is present. This weakest-link behavior is commonly described with extreme-value or Weibull-type statistics, although the choice of distribution is empirical~\cite{Gumbel1958,Weibull}. Within this framework, a stressed-area scaling relation was derived from a weakest-link statistical treatment of the measured breakdown-field distributions~\cite{Phan2021}. Gap spacing and stressed volume can also influence the breakdown, either through the field distribution itself or through contaminants in the liquid volume between the electrodes~\cite{Gerhold1994,Hayakawa2014,Kawashima1974,Weber1956,Gerhold1998,Acciarri2014,Auger2016,Hayakawa1997,Goshima1995,Goshima1995scale}. These effects are not always independent: changing the gap can change the stressed area, and changing the electrode shape can change both the maximum field and the distribution of possible initiation sites~\cite{Phan2021}.

For these reasons, comparisons of breakdown strength across cryogenic liquids are most informative when electrode material, surface finish, gap spacing, stressed area, pressure history, and measurement procedure are held fixed. The measurements reported here adopt this strategy so that the observed differences between LN$_2$, LHe, and LAr can be interpreted primarily in terms of liquid-dependent charge transport, thermodynamic stability against bubble formation, and surface-state evolution.


\section{Experimental apparatus, procedure, and statistical treatment}\label{sec:setup}

\subsection{Apparatus}\label{sec:apparatus}

\begin{figure}
    \centering
	\subfloat[Insert with Central Volume]{\includegraphics[width=0.40\linewidth, valign=c]{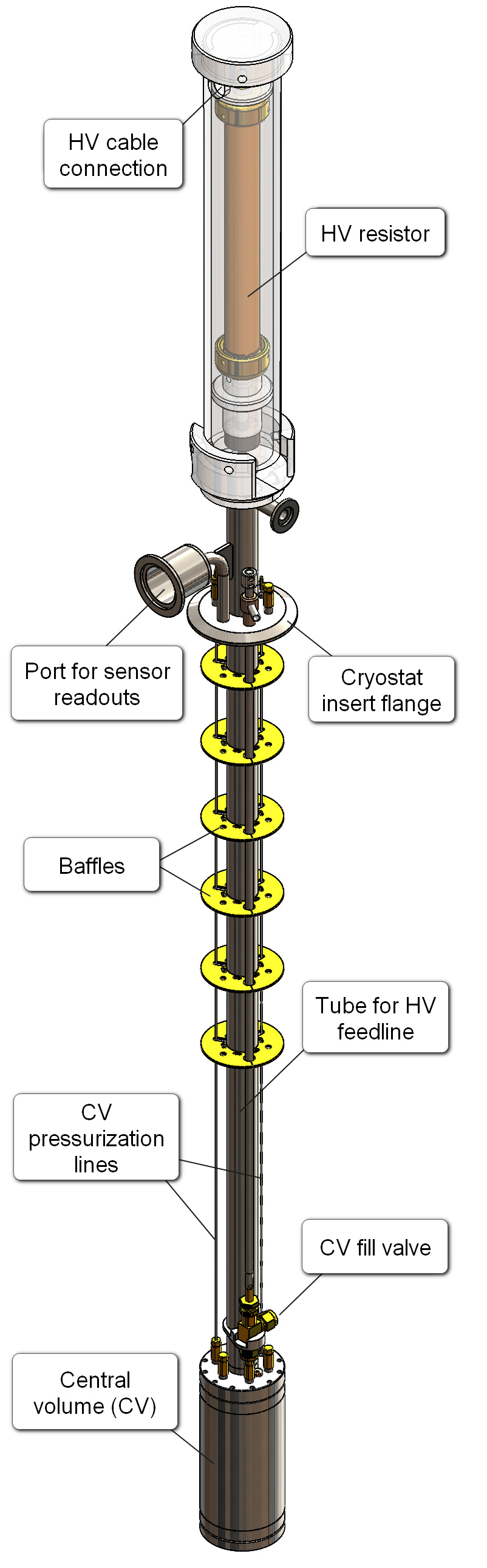}\label{fig:SSHV-diagram}}
	\subfloat[Central Volume]{\includegraphics[width=0.60\linewidth, valign=c]{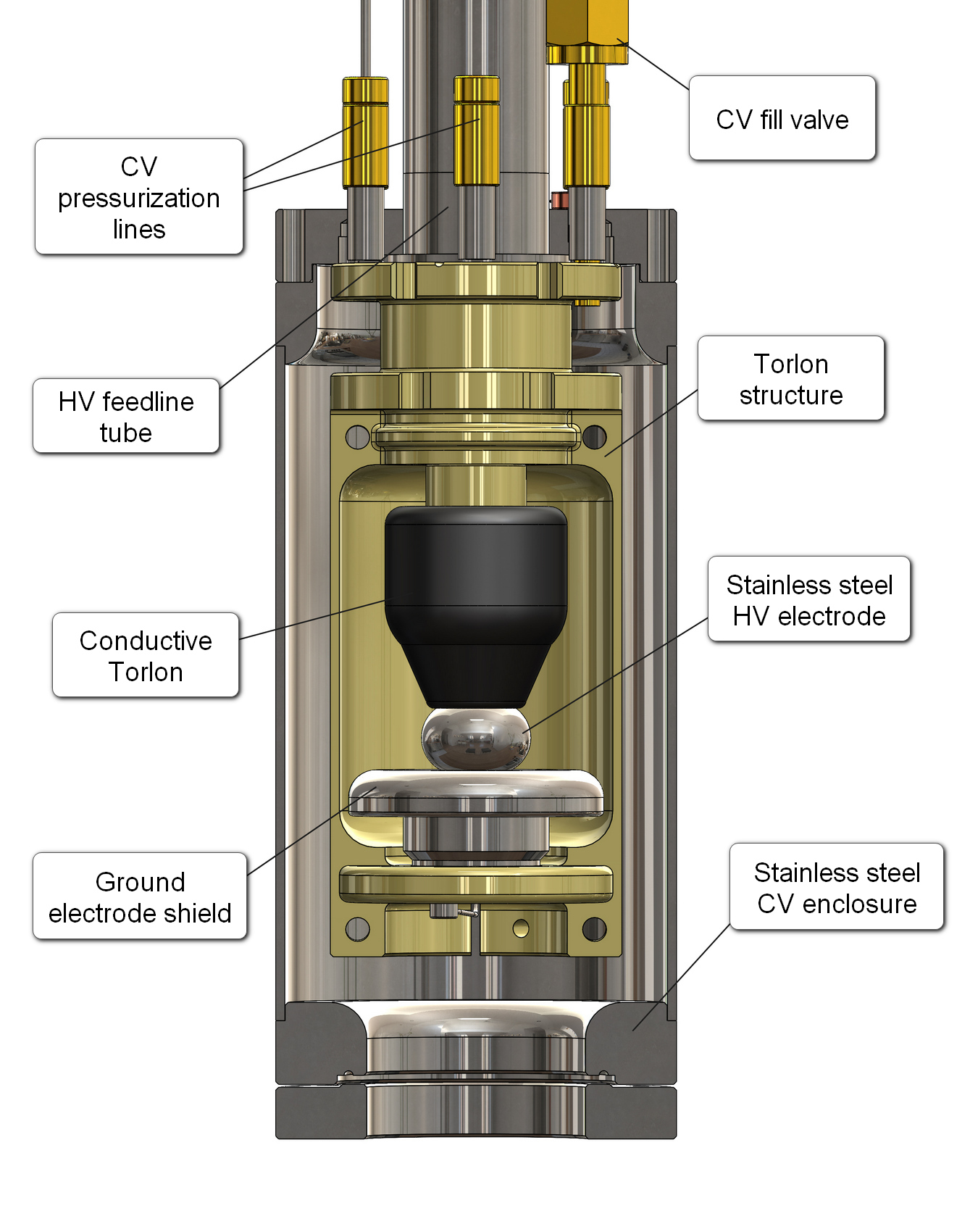}\label{fig:CV-diagram}} \\
    \caption{Experimental insert and central volume (CV) assembly. Panel~(a) shows the cryostat insert containing the CV, HV resistor, baffles, fill valve, and CV pressurization lines. Panel~(b) shows the CV containing the stainless-steel electrode pair, Torlon support structure, conductive Torlon connection to the HV feedline, fill valve, ground shield, and pressurization lines. The CV is filled with the target cryogenic liquid and can be pressurized independently of the surrounding dewar volume (DV).}
\label{fig:apparatus}
    \label{fig:SSHV-apparatus}
\end{figure}

The experimental setup used for the measurements in this work consisted of two primary components: a Janis Model 10 CNDT Research Dewar (cryostat) and an insert housing the electrodes (Figure~\ref{fig:SSHV-diagram}). Additional details of this apparatus, which was previously used in the study reported by Phan~\textit{et al.}~\cite{Phan2021}, can be found in that publication.  The electrode enclosure, referred to as the "central volume" (CV) and shown in Figure~\ref{fig:CV-diagram}, was a stainless-steel can independent of the internal "dewar volume" (DV) of the Janis Dewar. The CV was placed inside the DV. After the DV was filled with the target liquid cryogen, the liquid was introduced into the CV via a fill valve.

To further cool the liquid in both volumes, an evaporative cooling process was employed by pumping on the DV. For measurements at saturated vapor pressure, the CV fill valve was left open to the DV so
that the liquid in the CV remained at the saturated vapor pressure. For pressurizing the CV, the fill valve was closed and an external source of ultra-high-purity (99.999\%) gas was connected through a small-diameter tube and adjusted to the desired pressure. This method allowed for independent control of both temperature ($T$) and pressure ($P$). 

For each cryogen, the pressurization gas was the same species as the liquid under study: nitrogen gas for \LNtwo, helium gas for \LHe, and argon gas for \LAr. The gas was introduced through the CV pressurization line, and no in situ impurity monitor was located at the electrode gap. Consequently, the purity of the liquid in the high-field region is linked to the combined source purity, the CV volume, and any impurities released by breakdown events.

Unless otherwise stated, all pressures reported in this paper are absolute pressures measured for the CV. The local atmospheric pressure during these measurements was approximately \SI{590}{Torr} (the elevation of Los Alamos, New Mexico, is 2200~m); therefore, the pressurized conditions listed in Table~\ref{tab:runs} correspond to modest overpressures relative to the local atmosphere. The CV pressure was measured using a pressure gauge mounted on the CV gas-handling manifold. During each breakdown sequence the valve configuration between the CV and the pressure gauge was fixed, and pressure drops between the manifold gauge and the CV were negligible on the voltage-ramp time scale. The pressure values reported in Table~\ref{tab:runs} are the mean pressures during each breakdown sequence. The pressure variation within a sequence was typically less than \SI{1}{Torr}, and the gauge pressure uncertainty was \SI{0.5}{Torr}.

The liquid temperature was monitored using a combination of silicon diode and ruthenium oxide (Rox) temperature sensors located inside the DV and CV, respectively. The temperatures listed in Table~\ref{tab:runs} characterize the liquid bath near the saturated-vapor-pressure condition for each cryogen and are consistent with the corresponding lower-pressure CV measurements within the sensor uncertainty of 0.5~K at 77~K for the diode and 125~mK at 4.2~K for the Rox. The temperature changed negligibly during subsequent CV pressurization.
For \LHe operation, the system could be operated at temperatures as low as \SI{1.7}{K} and at CV pressures ranging from the saturated vapor pressure to approximately \SI{776}{Torr} absolute, corresponding to an overpressure of approximately \SI{186}{Torr} relative to the local ambient pressure.

The CV enclosure, shown in Figure~\ref{fig:CV-diagram}, contained stainless-steel electrodes separated by an average cold gap of approximately \SI{0.336}{mm}. The warm gap $d_w$ was measured with a Gapman\textsuperscript{\textregistered}
Gen3 Portable Electronic Feeler Gauge. The corresponding cold gap $d_c$ was then calculated by applying the thermal contraction of the stainless-steel electrode assembly and Torlon support structure between 292~K and the
operating temperature. The values of $d_{\mathrm{w}}$ and $d_{\mathrm{c}}$ are listed in Table~\ref{tab:runs}. The uncertainty in the warm-gap measurement was 0.010~mm, while the cold-gap uncertainty is dominated by an estimated $\pm5\%$ uncertainty in the thermal-contraction coefficients~\cite{Corruccini1961, Mann1977, Ventura1999}. The total uncertainty in the
cold gap is approximately 4\%.

\begin{figure}
    \centering
	\subfloat[Electrode field intensity map]{\includegraphics[width=0.510\linewidth]{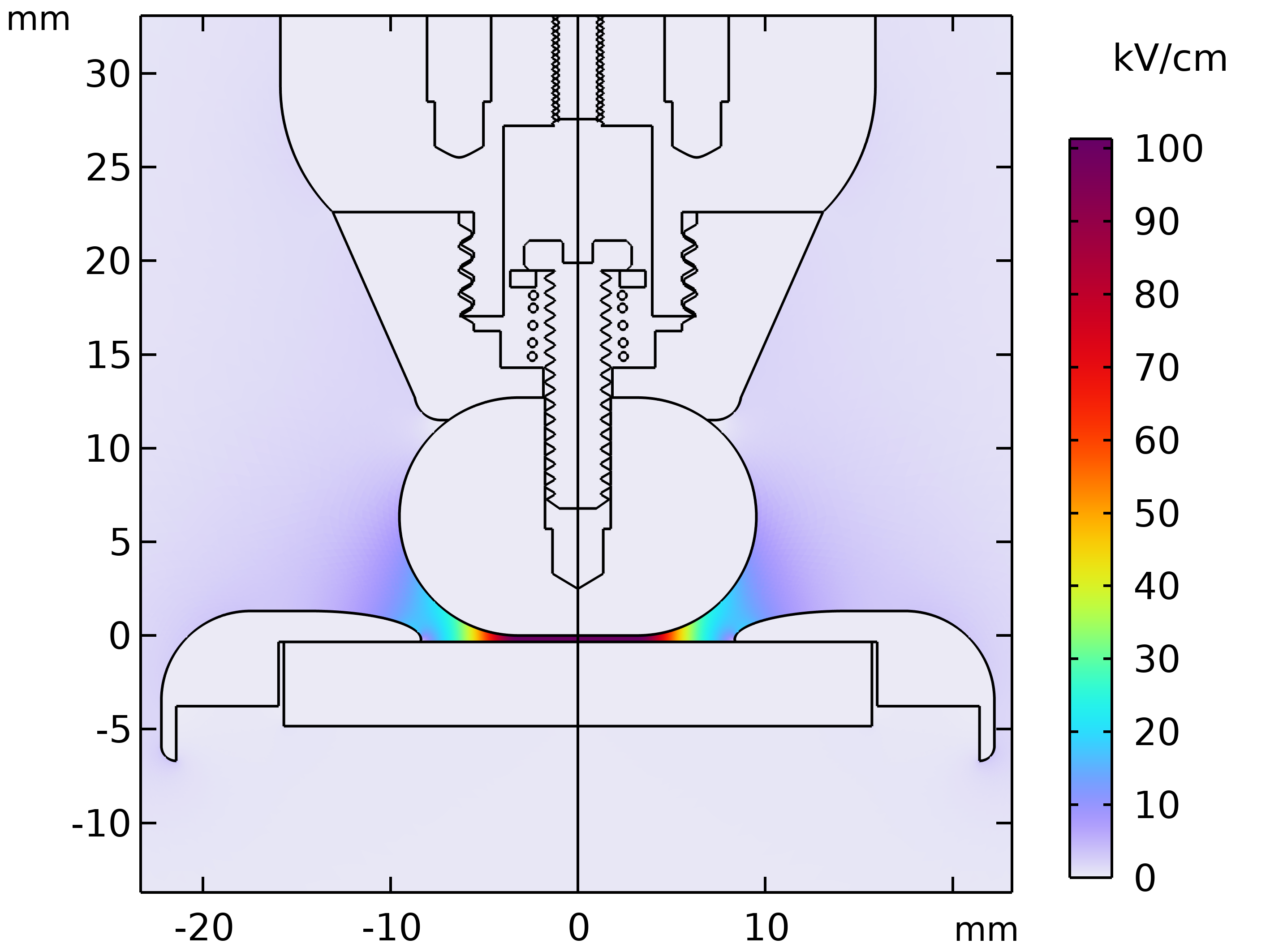}\label{fig:sshv-assem-field-map}}
	\subfloat[Field profile along cathode diameter]{\includegraphics[width=0.490\linewidth]{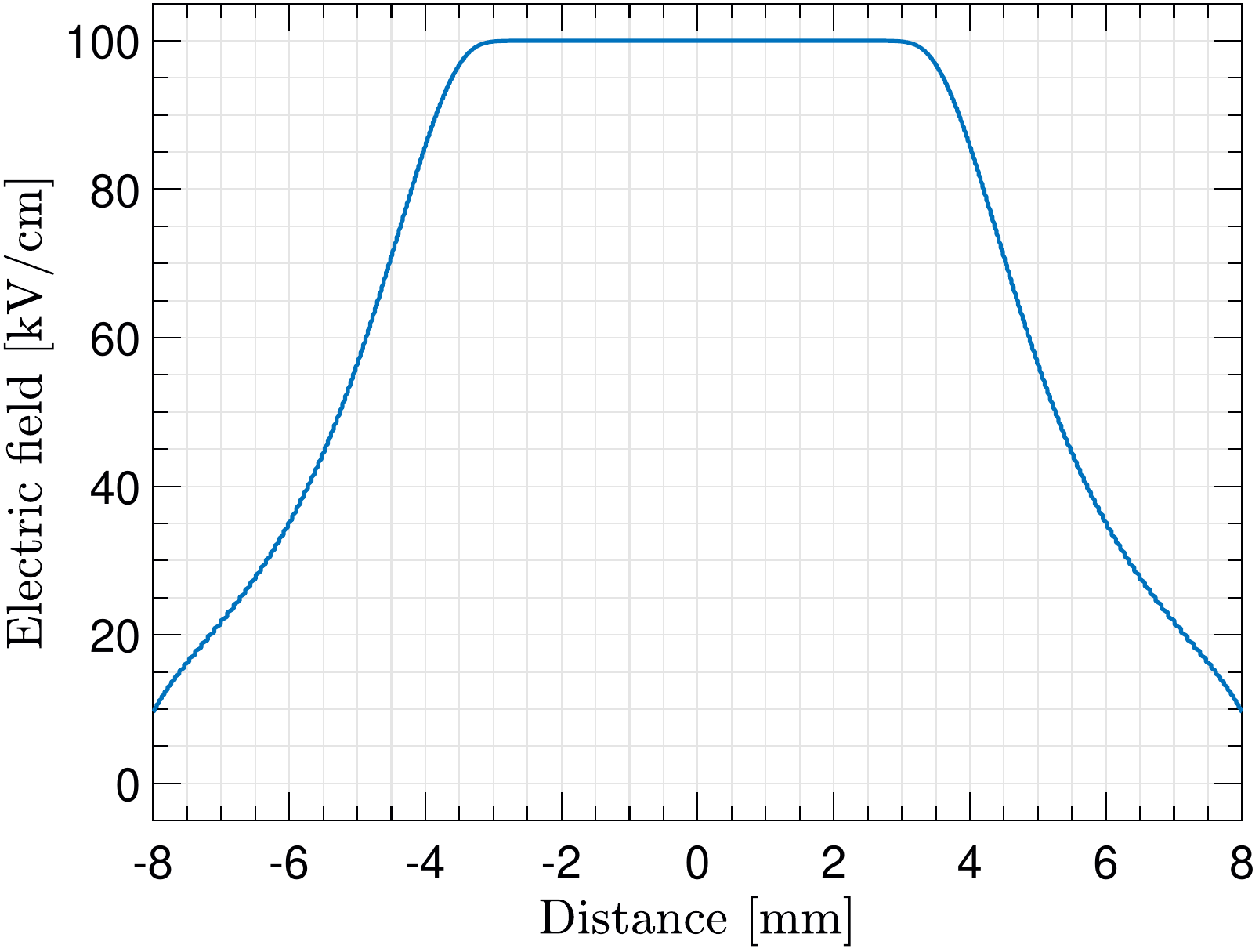}\label{fig:sshv-assem-field-vs-dist}} 
    \caption{COMSOL electrostatic calculation of the field between the electrodes. Panel (a) shows the two-dimensional field map with the field scaled so that the maximum field is \(100~\mathrm{kV\,cm^{-1}}\). Panel (b) shows the
corresponding field profile along the diameter of the grounded electrode. The electrostatic solution scales linearly with applied voltage; therefore, the breakdown fields reported in the text are
calculated event-by-event as \(E_{\mathrm{b}}=V_{\mathrm{b}}/d_{\mathrm{c}}\), using the cold gap listed in Table~\ref{tab:runs}. The stressed area used in this work
is defined by the region in which the local field exceeds 70\% of the maximum field. The plotted field is the local COMSOL field normalized for visualization, whereas the breakdown fields used for
liquid-to-liquid comparisons are the nominal values \(E_{\mathrm{b}}=V_{\mathrm{b}}/d_{\mathrm{c}}\).}
    \label{fig:field-map}
\end{figure}

The electrostatic field distribution calculated using COMSOL is shown in Figure~\ref{fig:field-map}. Figure~\ref{fig:field-map}a
shows the two-dimensional field-intensity map between the electrodes,
while Figure~\ref{fig:field-map}b shows the corresponding field profile along the diameter of the grounded electrode. The stressed area, $S_0$,
was determined from this calculation as the electrode area over which the local field exceeded 70\% of the maximum field. Under this convention,
\begin{equation}
S_0 = 0.66~\mathrm{cm^2}.
\label{eq:stressed-area}
\end{equation}
The 70\% threshold is used consistently throughout this work. Because the same electrode geometry is used for all liquids, the cross-liquid comparisons are insensitive to the absolute stressed-area convention, although comparisons with other experiments depend on this definition.


\subsection{Procedure}\label{sec:procedure}
Once the CV reached the desired temperature and pressure, breakdown measurements were performed with the following repeated voltage-ramp sequence. A GAMMA RR-series HV power supply (HVPS) was
connected to the upper (oblate) electrode through an external 8~$\mathrm{M\Omega}$ series resistor (the HV resistor in Figure~\ref{fig:SSHV-apparatus}a), the HV feedline, and the conductive Torlon element inside the CV. The conductive element was fabricated from Torlon~4435, and independent cryogenic testing confirmed that it retained electrical continuity over the full temperature range used in the present
measurements. The lower electrode was held at ground potential and connected to the current-trigger circuit. The HV electrode was held at positive polarity relative to ground for all measurements; polarity was not varied. With this convention, the grounded lower electrode served as the cathode for field-emission processes in the liquid gap. Unless otherwise stated, the voltage was ramped at a constant rate of \SI{500}{V.s^{-1}} until either breakdown occurred or the programmed maximum voltage, typically \SI{40}{kV}, was reached.

The breakdown voltage $V_\mathrm{b}$ is defined as the HVPS voltage-monitor value at the time the current-trigger signal crossed its threshold. Before the current transient, leakage currents were small, so the voltage drop across the external series resistor was negligible compared with the voltage calibration uncertainty of 2\%. Breakdown was identified by a rapid current transient on the grounded electrode. The trigger threshold was \SI{0.1}{A}, chosen to be well above the leakage-current baseline and below the current amplitude of all observed breakdown events. When triggered, the interlock disabled the HVPS output and forced the voltage to zero. The trigger-to-interlock delay was \SI{0.5}{ms}, primarily set by the relay-switch operating time, corresponding to a voltage-registration uncertainty of \SI{0.25}{V} at a ramp rate of \SI{500}{V.s^{-1}}. This timing contribution is negligible compared with the voltage calibration uncertainty. No ramps in the datasets reported here reached the programmed maximum voltage without breakdown.  Therefore, no right-censoring correction was applied.

The energy associated with each breakdown was not measured on an event-by-event basis. Before breakdown, the electrostatic energy stored in the
HV circuit can be estimated as
\begin{equation}
    U_{\mathrm{stored}}(\Vb)
    =
    \frac{1}{2} C_{\mathrm{HV}} \Vb^2 ,
    \label{eq:stored-energy}
\end{equation}
where \(C_{\mathrm{HV}}\) is the effective capacitance of the HV
feedthrough, cable, electrode assembly, and power-supply output. The estimated
total capacitance was \(C_{\mathrm{HV}}\simeq\SI{85}{pF}\), corresponding to
\(U_{\mathrm{stored}}\simeq\SI{68}{mJ}\) at \(\Vb=\SI{40}{kV}\). The
electrode-to-electrode capacitance in the CV was much smaller,
approximately \(C_{\mathrm{gap}}\simeq\SI{3}{pF}\), corresponding to a local
gap electrostatic energy of approximately \(\SI{2.4}{mJ}\) at
\(\SI{40}{kV}\). For an event at breakdown voltage \(\Vb\), both estimates
scale as \((\Vb/\SI{40}{kV})^2\).  The capacitance estimates were obtained from direct capacitance measurements and electrostatic modeling.

We note that the larger value is the energy initially
stored in the external HV circuit, not the energy necessarily deposited in the liquid gap. Because most of the capacitance was associated with the cable, feedthrough, and power-supply side of the circuit, the \SI{8}{M\ohm} series resistor dissipated much of this energy. Much of the energy stored in the capacitance between the series resistor and the HV electrode was absorbed by the conductive Torlon 4435 element (which also serves as the electrode holder).

After each breakdown, the HVPS output was held at zero for \SI{30}{s} before
the next ramp. This recovery interval allowed the HVPS and interlock circuit
to reset and provided time for short-lived electrical and thermal disturbances
near the electrode gap to decay. The recovery time was fixed for all datasets
and was not varied.

After completing the measurements for a given cryogen, the system was allowed to warm passively to room temperature. The DV, CV, and gas-handling volumes were then evacuated to below \SI{1e-3}{Torr}, with pumping sustained for at least \SI{24}{h}. The insert, CV, and electrodes remained installed and mechanically undisturbed throughout the full sequence of runs. This procedure preserved the electrode geometry, gap spacing, material, and surface finish as much as possible, while allowing the cryogenic liquid to be changed between runs. The chronological sequence was \LNtwo, \LHe, \LAr, and \LNtwo. Within each cryogen run, the lower-pressure dataset was acquired before the
pressurized dataset. Several hours elapsed between completion of the low-pressure sequence and the start of the pressurized sequence because of the
time required to pressurize the CV. Pressure-dependent comparisons within a cryogen
may also include any conditioning accumulated during the lower-pressure
sequence. However, no clear conditioning trend was observed in these data. The final \LNtwo run provides an empirical check on run-order and electrode-surface evolution.

Breakdown fields are reported as
\begin{equation}
    \Eb = \frac{\Vb}{d_{\mathrm{c}}},
    \label{eq:field-def}
\end{equation}
where \(\Vb\) is the measured breakdown voltage and \(d_{\mathrm{c}}\) is the cold electrode spacing listed in Table~\ref{tab:runs}. Here \(E_{\mathrm{b}}\) is a nominal gap-averaged breakdown field used for
cross-liquid comparison. It is not the local maximum field from the COMSOL
calculation. Because the same electrode geometry is used throughout the
measurement sequence, this nominal field definition provides a consistent
reference for comparing the cryogens. The same cold-gap correction procedure is applied to all field comparisons.  The uncertainty in $\Eb$ is obtained from the voltage and gap uncertainties,
\begin{equation}
    \left(\frac{\sigma_{\Eb}}{\Eb}\right)^2
    =
    \left(\frac{\sigma_{\Vb}}{\Vb}\right)^2
    +
    \left(\frac{\sigma_{d_{\mathrm{c}}}}{d_{\mathrm{c}}}\right)^2 ,
    \label{eq:field-uncertainty}
\end{equation}
where $\sigma_{\Vb}$ includes voltage calibration and timing contributions, and $\sigma_{d_{\mathrm{c}}}$ is the estimated cold-gap uncertainty. Because the same electrodes and gap model are used for all liquids, uncertainty in $d_{\mathrm{c}}$ affects the absolute field scale rather than the relative comparison between cryogens.

\begin{table}[t]
\centering
\caption{
Experimental run conditions. 
The run number gives the chronological order of the measurements. 
The temperature $T$ was measured near the saturated-vapor-pressure condition for each cryogen. 
$P_1$ and $P_2$ are the average absolute CV pressures for the two pressure settings in each run. 
$d_{\mathrm{w}}$ and $d_{\mathrm{c}}$ are the warm and cold electrode spacings, respectively.
}
\label{tab:runs}
\begin{tabular}{clccccc}
\toprule
Run & Liquid & $T$ [K] & $P_1$ [Torr] & $P_2$ [Torr] & $d_{\mathrm{w}}$ [mm] & $d_{\mathrm{c}}$ [mm] \\
\midrule
1 & Nitrogen & 75.6 & 600 & 760 & 0.371 & 0.338 \\
2 & Helium   & 3.99  & 602 & 768 & 0.371 & 0.328 \\
3 & Argon    & 85.3  & 612 & 760 & 0.371 & 0.341 \\
4 & Nitrogen & 75.6 & 602 & 760 & 0.371 & 0.338 \\
\bottomrule
\end{tabular}
\end{table}

\subsection{Statistical treatment of breakdown distributions}
\label{sec:statistics}

Electrical breakdown in cryogenic liquids is stochastic.  Therefore, each
thermodynamic condition is characterized by the empirical distribution of
breakdown voltages rather than by a single characteristic value. For each
pressure setting, an event is defined as one voltage ramp ending in a detected
breakdown. The breakdown voltage \(\Vb\) is converted to a gap-corrected
breakdown field using Eq.~\eqref{eq:field-def}. Unless otherwise stated, quoted
central values are arithmetic means of the observed
breakdown voltages for a given liquid and pressure.

For each dataset, we report the number of breakdowns \(N\), the
arithmetic mean breakdown voltage \(\langle V_{\mathrm{b}}\rangle\), the standard
deviation \(\sigma_{V_{\mathrm{b}}}\), the median, and the empirical 10th and 90th
percentiles, denoted \(Q_{10}(V_{\mathrm{b}})\) and \(Q_{90}(V_{\mathrm{b}})\), respectively.
The quantity \(Q_{10}(V_{\mathrm{b}})\) is the breakdown voltage below which
10\% of the observed events occur, whereas \(Q_{90}(V_{\mathrm{b}})\) is the
breakdown voltage below which 90\% of the observed events occur. Consequently, the interval $\left[Q_{10}(V_{\mathrm{b}}),\,Q_{90}(V_{\mathrm{b}})\right]$ contains the central 80\% of the observed breakdown-voltage distribution. These percentiles, like \(\sigma_{V_{\mathrm{b}}}\), characterize
the event-to-event spread of the measured distribution and should not
be interpreted as confidence limits or uncertainties on the mean.

The full-distribution summary statistics are compiled in
Table~\ref{tab:distribution-statistics}. The principal values used in the
subsequent cross-liquid comparisons are collected separately in
Table~\ref{tab:principal-values}. These include the component
means for the bimodal 602~Torr LHe dataset and the first-breakdown and
approximate plateau values for the non-stationary 612~Torr LAr dataset.
The dominant systematic uncertainties relevant to the cross-liquid
field comparison are discussed in
Section~\ref{sec:liquid-comparison}.

The time-series plots are used to identify non-stationary behavior before
interpreting the corresponding histograms. In particular, the \SI{612}{Torr}
\LAr dataset shows a turn-on feature in which the breakdown voltage evolves
with event number. For this dataset, the first breakdown and the approximate
plateau value are treated as descriptive quantities rather than as statistics
of a stationary distribution.

For datasets that show visually distinct lower- and higher-field populations,
such as the \SI{602}{Torr} \LHe dataset, component means are reported using an
empirical voltage threshold corresponding to the valley between the histogram
peaks. These component assignments are used to compare the observed populations. A more detailed empirical cumulative-hazard
comparison for the \LHe components is given in Section~\ref{sec:lhe-bimodal}.


\section{Results}\label{sec:results}

\subsection{Liquid nitrogen}\label{sec:LN2-Run1}

\begin{figure}
    \centering
	\subfloat[LN$_2$, 600 Torr]{\includegraphics[width=0.501\linewidth]{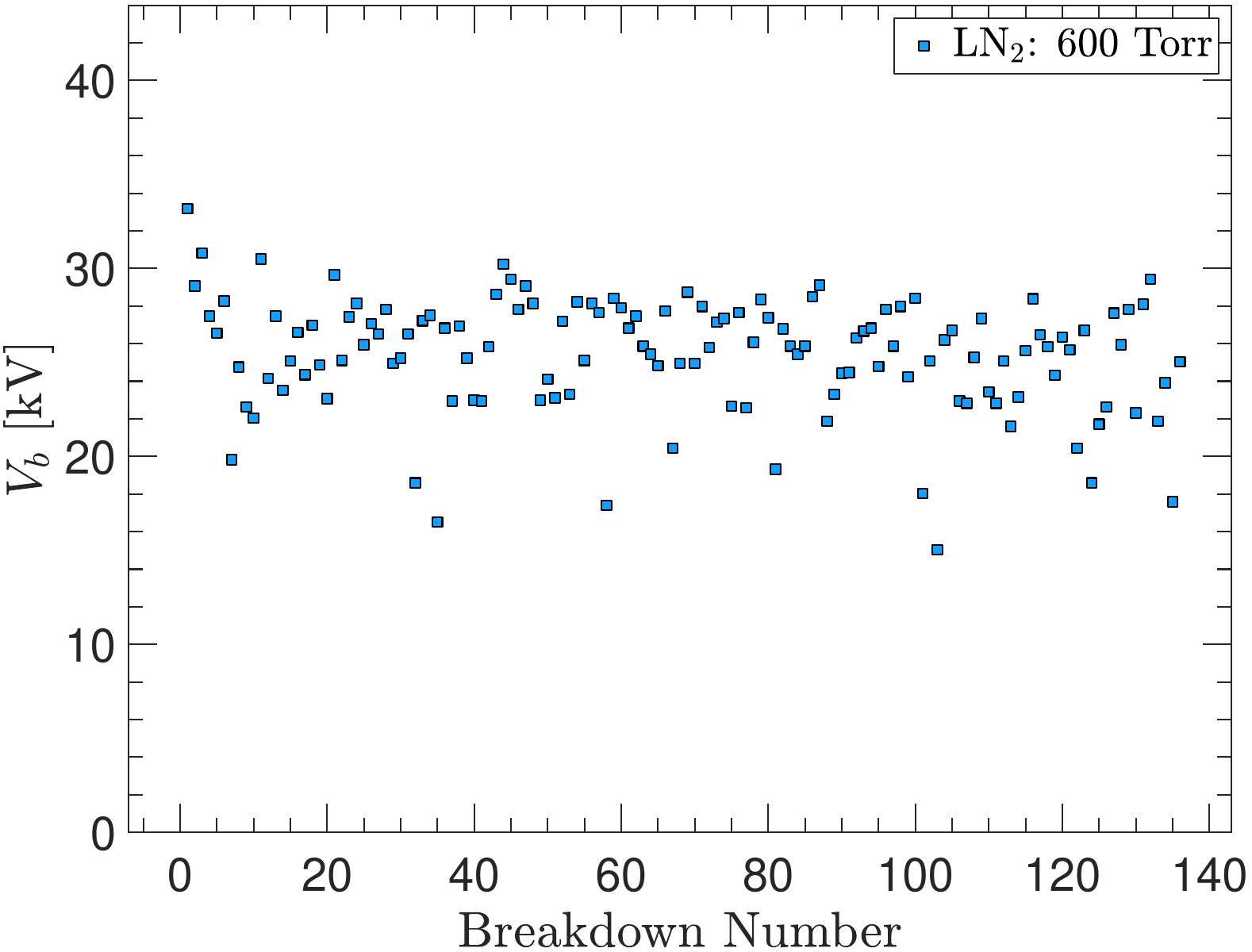}\label{fig:LN2-Run1-seq-600Torr}}
	\subfloat[LN$_2$, 600 Torr]{\includegraphics[width=0.499\linewidth]{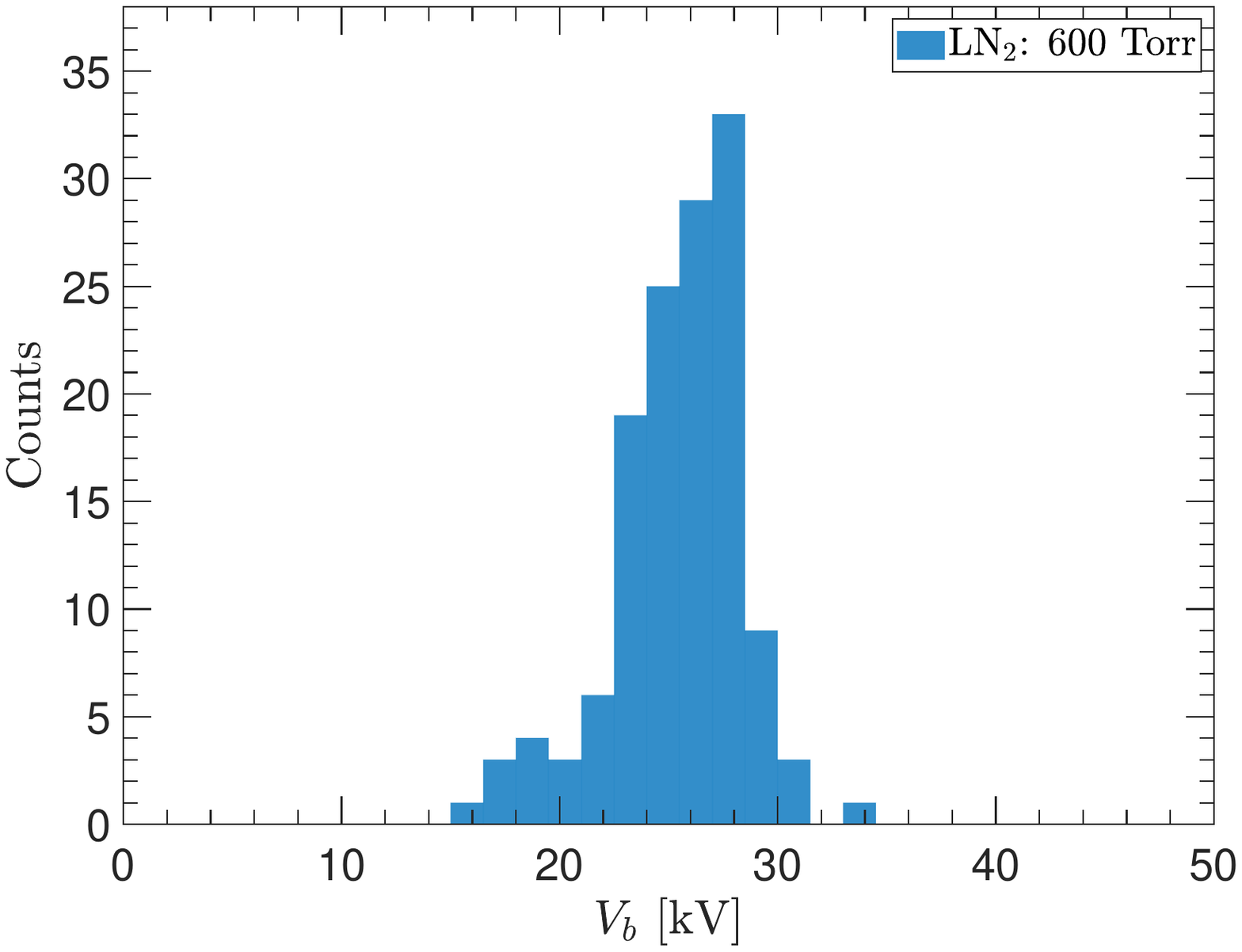}\label{fig:LN2-Run1-hist-600Torr}} \\
	\subfloat[LN$_2$, 760 Torr]{\includegraphics[width=0.498\linewidth]{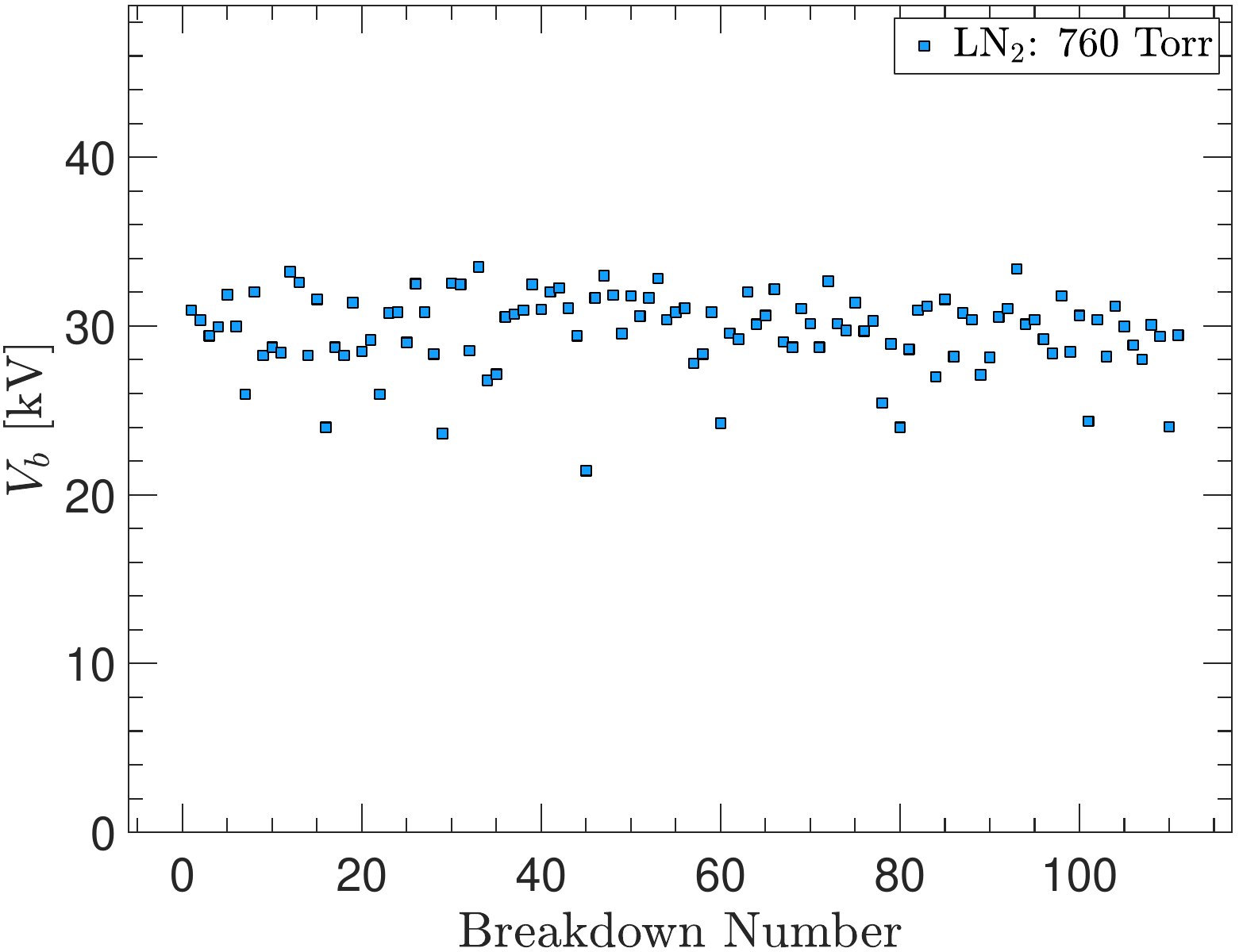}\label{fig:LN2-Run1-seq-760Torr}}
    \subfloat[LN$_2$, 760 Torr]{\includegraphics[width=0.502\linewidth]{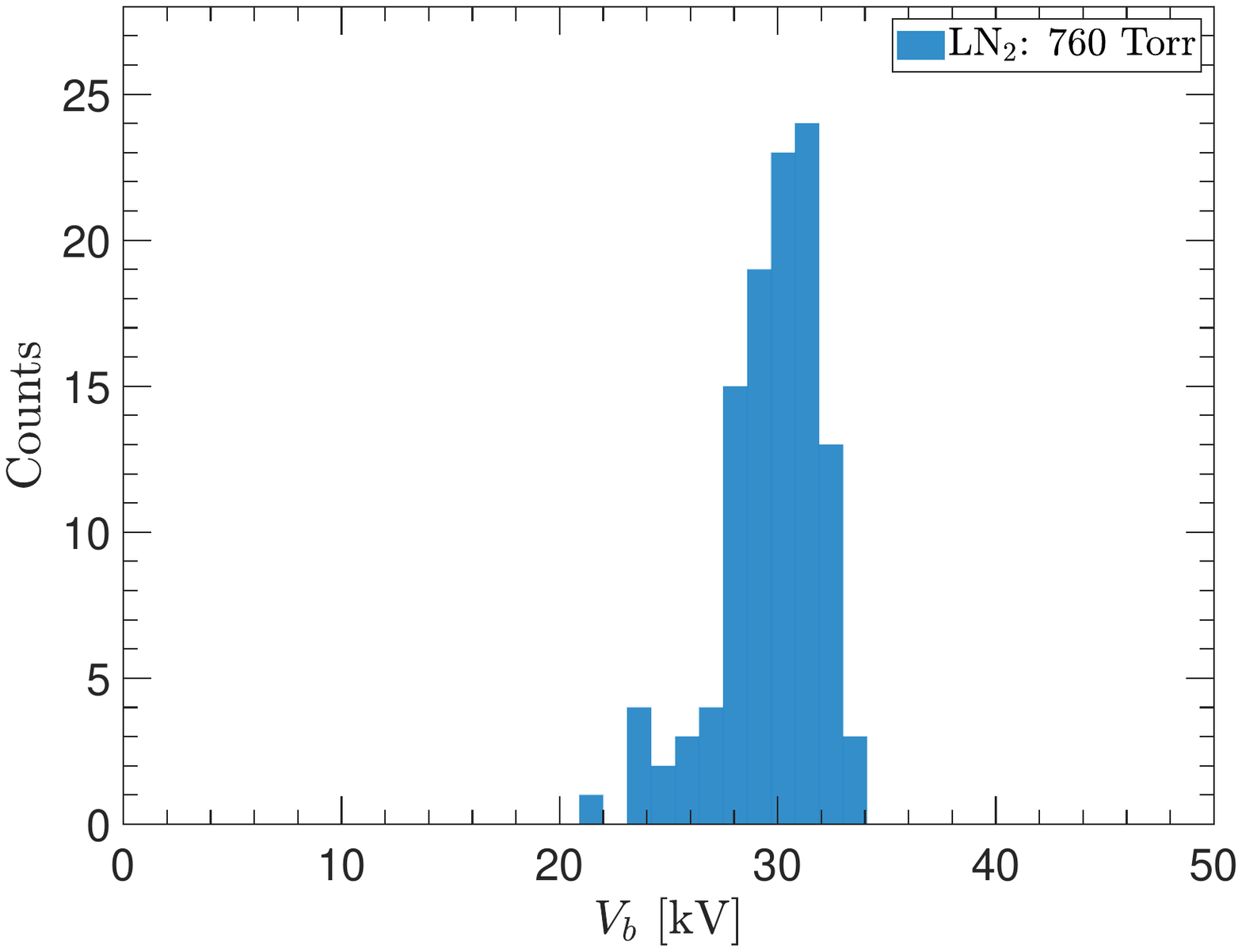}\label{fig:LN2-Run1-hist-760Torr}}
    \caption{Breakdown-voltage time series and histograms for \LNtwo run~1. Panels~(a,b) show the \SI{600}{Torr} dataset with $N=136$ breakdowns; panels~(c,d) show the \SI{760}{Torr} dataset with $N=111$ breakdowns. The mean breakdown voltages are \SI{25.4}{kV} at \SI{600}{Torr} and \SI{29.7}{kV} at \SI{760}{Torr}.}
\label{fig:ln2-run1}
\end{figure}

\begin{figure}
    \centering
	\subfloat[LN$_2$, 602 Torr]{\includegraphics[width=0.508\linewidth]{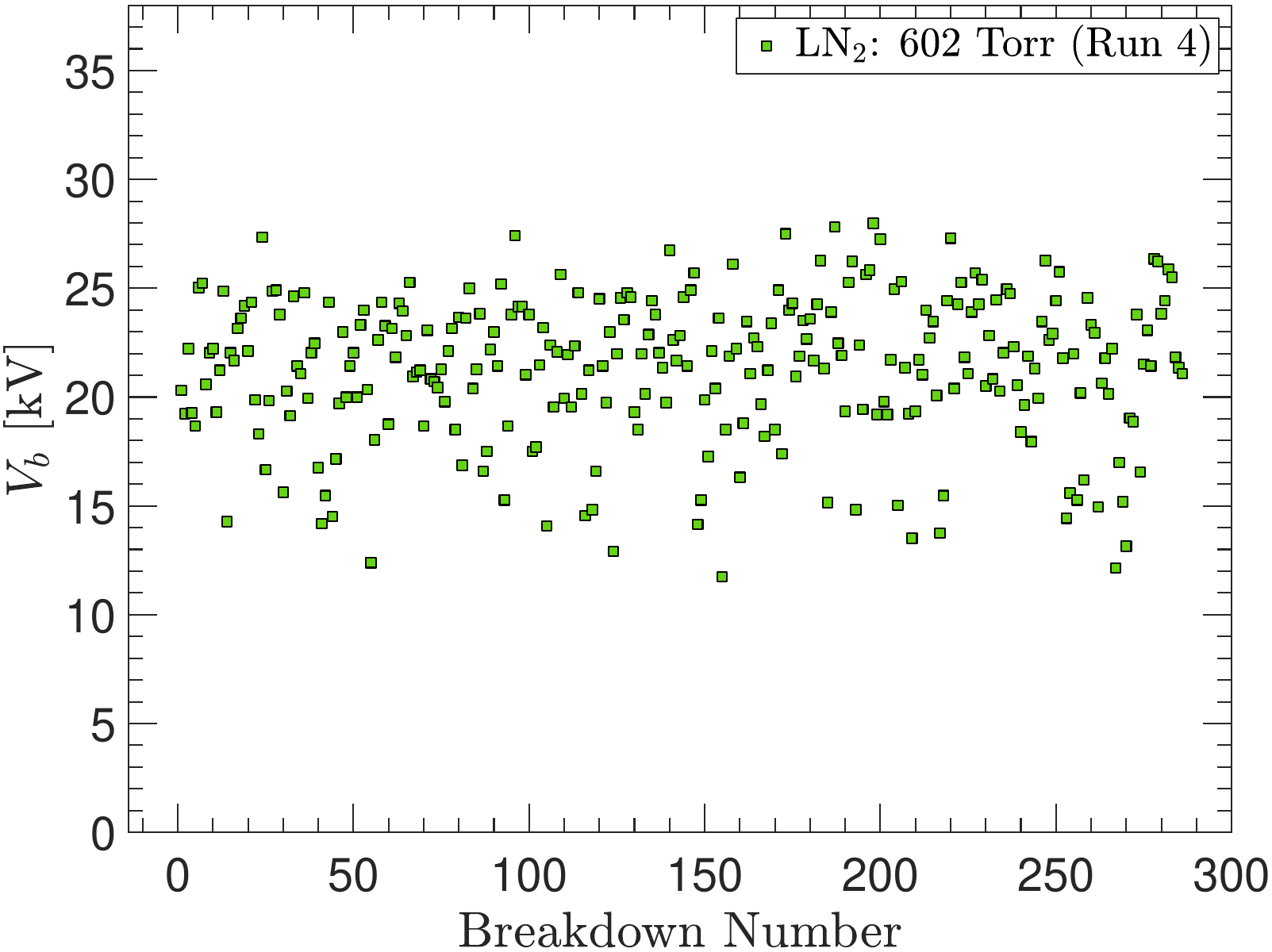}\label{fig:LN2-Run4-seq-602Torr}}
	\subfloat[LN$_2$, 602 Torr]{\includegraphics[width=0.492\linewidth]{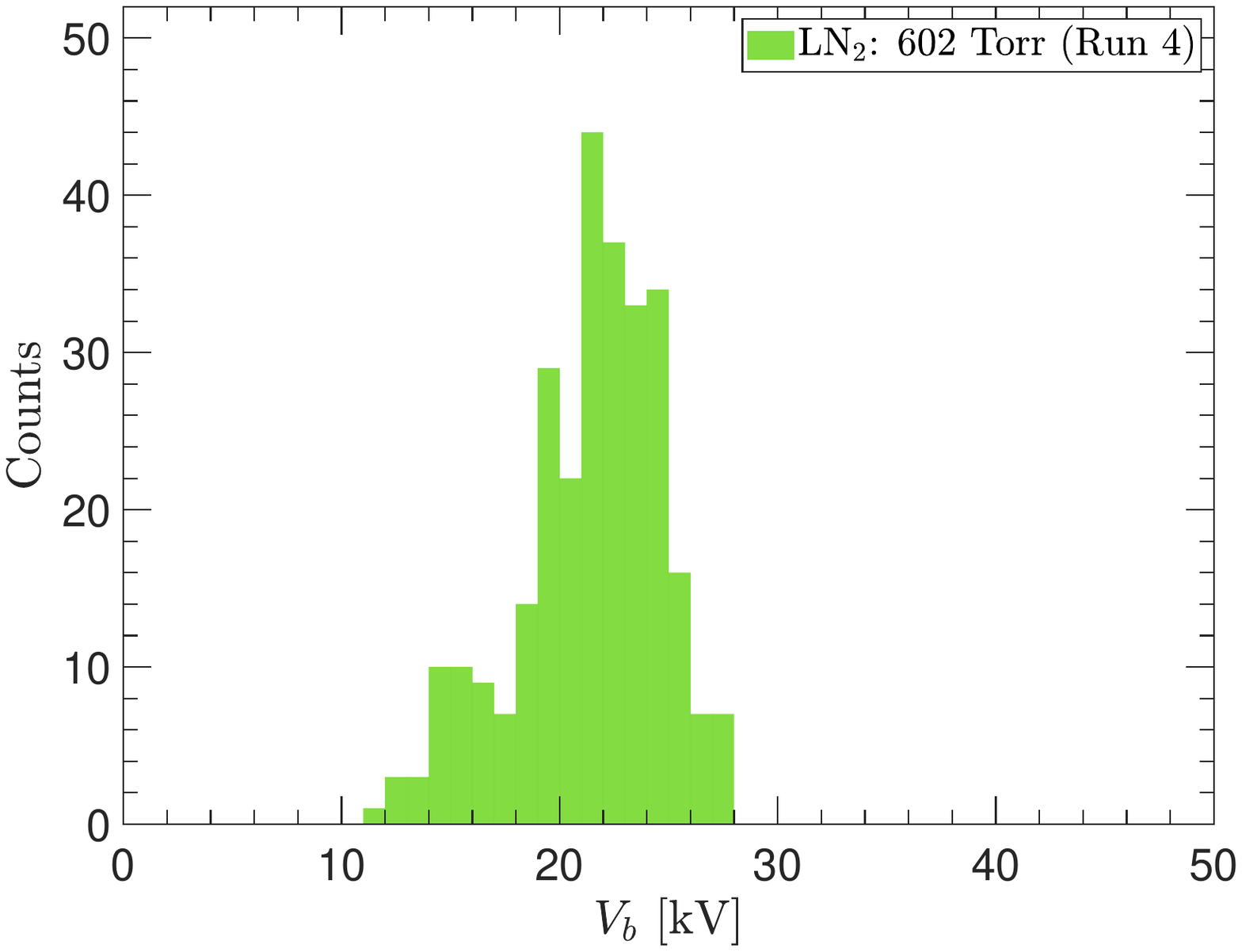}\label{fig:LN2-Run4-hist-602Torr}} \\
	\subfloat[LN$_2$, 760 Torr]{\includegraphics[width=0.504\linewidth]{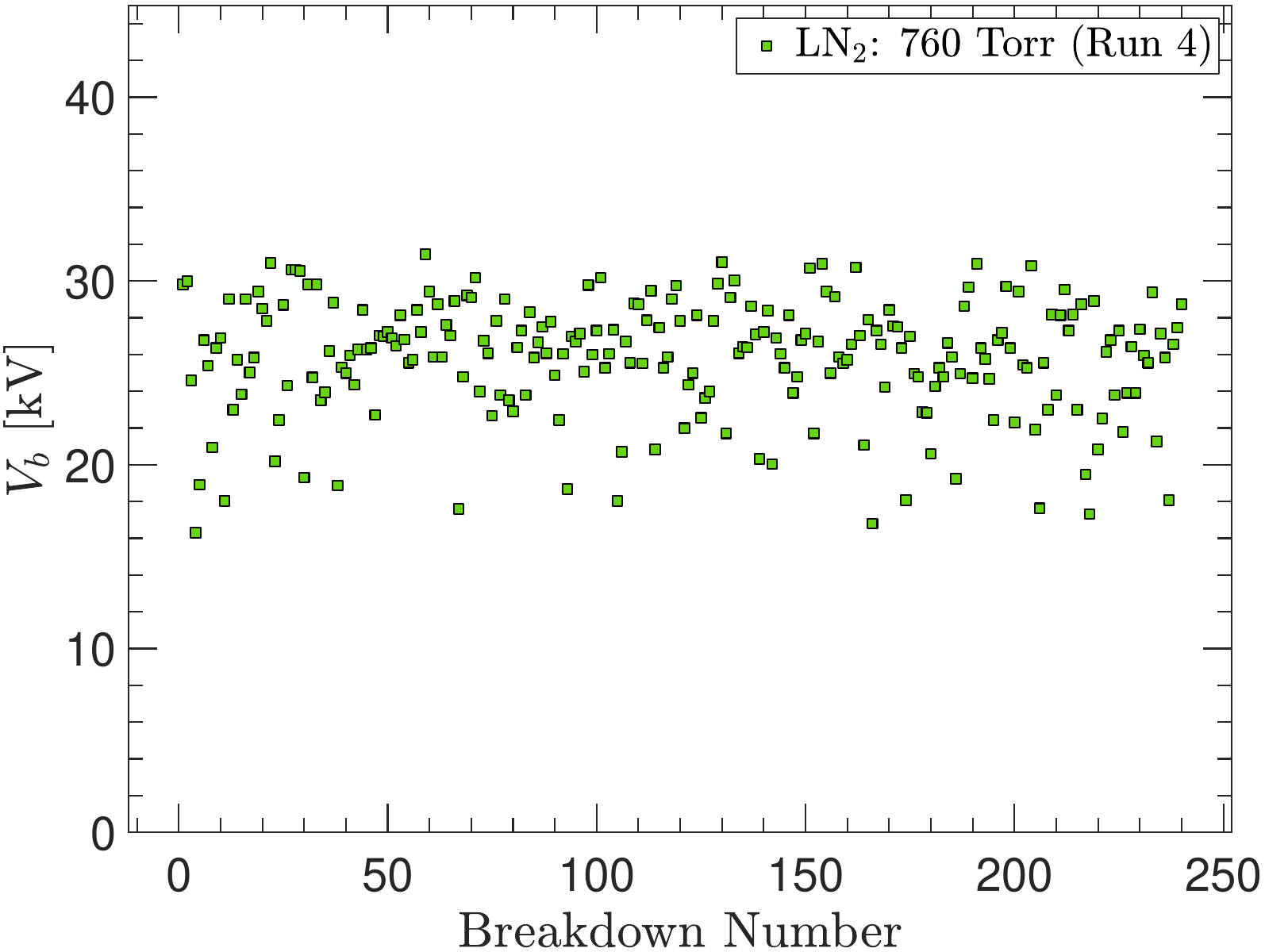}\label{fig:LN2-Run4-seq-760Torr}}
    \subfloat[LN$_2$, 760 Torr]{\includegraphics[width=0.496\linewidth]{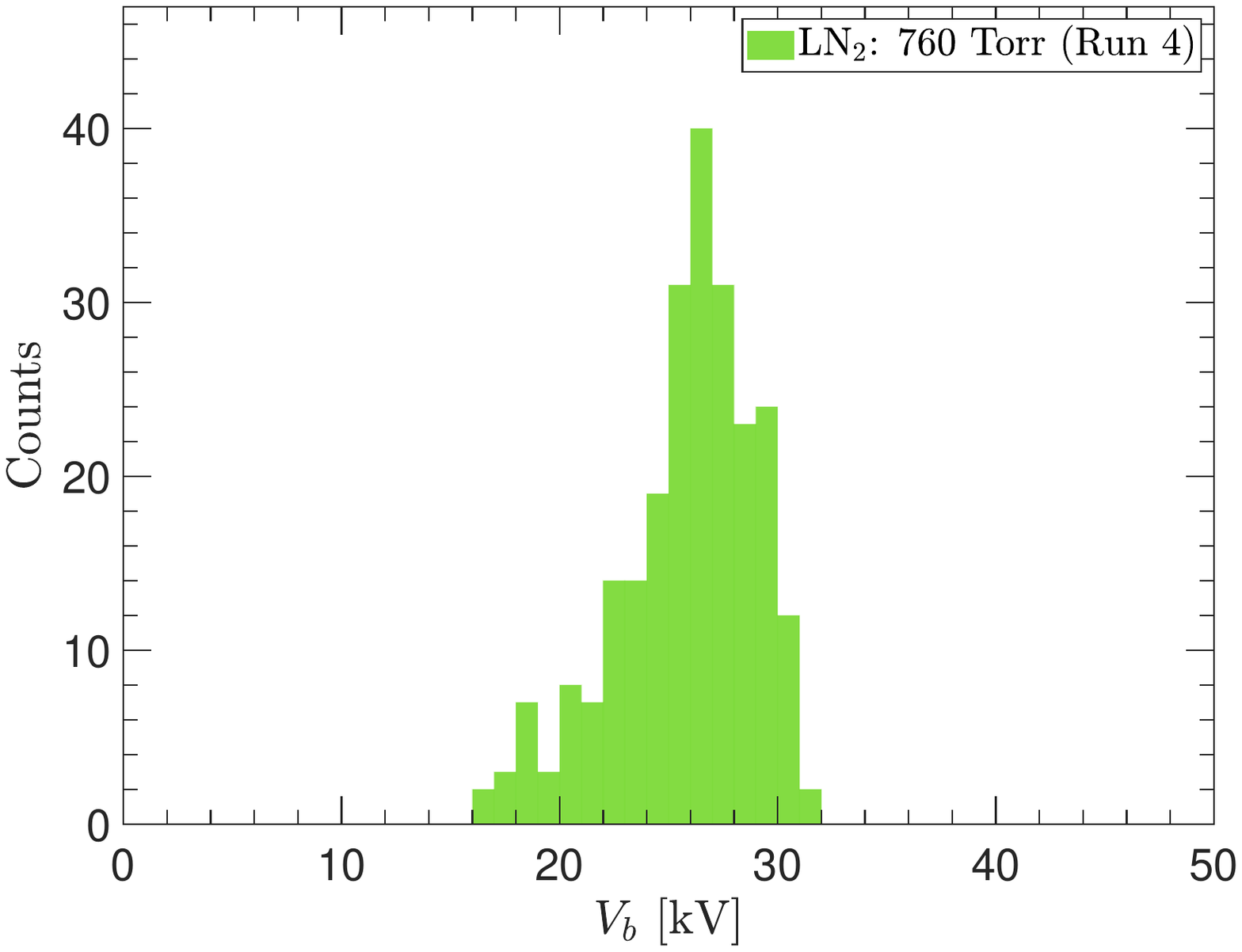}\label{fig:LN2-Run4-hist-760Torr}}
    \caption{Breakdown-voltage time series and histograms for \LNtwo run~4. Panels~(a,b) show the \SI{602}{Torr} dataset with $N=286$ breakdowns; panels~(c,d) show the \SI{760}{Torr} dataset with $N=240$ breakdowns. The mean breakdown voltages are \SI{21.4}{kV} at \SI{602}{Torr} and \SI{25.8}{kV} at \SI{760}{Torr}. The run-to-run shift relative to run~1 is used as an empirical estimate of
    the systematic variation associated with electrode surface state.}
    \label{fig:ln2-run4}
\end{figure}

The \LNtwo breakdown-voltage time series and distributions in
Figures~\ref{fig:ln2-run1} and~\ref{fig:ln2-run4} show that pressurization from
the saturated-vapor-pressure condition to approximately \SI{760}{Torr}
increases the measured breakdown voltage in both cooldowns. However, the
second \LNtwo cooldown, run~4, yielded lower breakdown voltages than run~1,
even though the apparatus was not mechanically reconfigured between the two
measurements. The electrodes experienced minimal air exposure during the
cryogenic sequence. Therefore, the difference between \LNtwo run~1
and run~4 is consistent with a change in electrode surface state
during the sequence of cooldowns and breakdown measurements. A possible contributor is modification of the electrode surface oxide: before the initial LN2
cooldown, the electrodes had been exposed to ambient air for several weeks.  Furthermore, repeated breakdowns may have changed microscopic asperities or altered adsorbed contaminants. Swan and Lewis~\cite{Swan1960} observed that more heavily oxidized stainless-steel electrodes produced higher breakdown fields in \LAr, with a difference comparable to the run-to-run shift observed here.  

Conventional electrical conditioning, in which repeated discharges deactivate prominent field-emission sites, would generally be expected to increase \(V_{\mathrm{b}}\) with event number and therefore produce a positive slope within an individual time series. By contrast, if the initially air-exposed oxide state was associated with higher breakdown strength, its modification during the intervening cooldowns and discharge sequences could produce a downward shift in the characteristic breakdown field between run~1 and run~4 without necessarily producing a negative slope within either individual dataset.  The absence of a clear positive trend in Figures~\ref{fig:ln2-run1} and~\ref{fig:ln2-run4} therefore provides no evidence for strong within-run electrical conditioning, but does not exclude slower cumulative evolution of the oxide, adsorbed contaminants, or other surface features between cooldowns.  In fact, a weak downward drift is visible
in the 600~Torr data of Figure~\ref{fig:ln2-run1}, with a less pronounced indication in the 760~Torr data, although both trends are small compared with the event-to-event scatter.

\begin{table}[t]
\centering
\caption{
Summary statistics for the measured breakdown-voltage distributions. The quantities \(Q_{10}(V_{\mathrm{b}})\) and \(Q_{90}(V_{\mathrm{b}})\) denote the empirical
10th and 90th percentiles of the breakdown voltage, respectively, such that the interval
\(\left[Q_{10}(V_{\mathrm{b}}),Q_{90}(V_{\mathrm{b}})\right]\) contains the central 80\% of the observed events. The standard deviation and percentile interval
describe the event-to-event spread of each distribution and are not uncertainties on the mean. For the non-stationary 612~Torr LAr dataset,
the tabulated values describe the complete event sequence; the first-breakdown and approximate plateau values used in the discussion
are reported separately in Table~\ref{tab:principal-values}.
}
\label{tab:distribution-statistics}
\begin{tabular}{llccccccc}
\toprule
Run &
Liquid &
\(P\) [Torr] &
\(N\) &
\(\langle V_{\mathrm{b}}\rangle\) [kV] &
\(\sigma_{V_{\mathrm{b}}}\) [kV] &
Median [kV] &
\(Q_{10}(V_{\mathrm{b}})\) [kV] &
\(Q_{90}(V_{\mathrm{b}})\) [kV]
\\
\midrule
1 & \LNtwo & 600 & 136 & 25.4 & 3.04 & 25.9 & 21.9 &  28.4 \\
1 & \LNtwo & 760 & 111 & 29.7 & 2.31 & 30.1 & 26.9 & 32.3 \\
2 & \LHe   & 602 & 200 & 18.5 & 5.17 & 20.2 & 11.3 & 24.1 \\
2 & \LHe   & 768 & 160 & 25.0 & 2.97 & 25.4 & 20.7 & 28.4 \\
3 & \LAr   & 612 & 222 & 14.2 & 2.54 & 14.7 & 10.8 & 17.3 \\
3 & \LAr   & 760 & 154 & 18.8 & 1.92 & 18.9 & 15.8& 21.4 \\
4 & \LNtwo & 602 & 286 & 21.4 & 3.33 & 21.8 & 16.3 & 25.2 \\
4 & \LNtwo & 760 & 240 & 25.8 & 3.16 & 26.4 & 21.2 & 29.5 \\
\bottomrule
\end{tabular}
\end{table}

\begin{table}[t]
\centering
\caption{
Principal breakdown values used for the cross-liquid comparisons. Pressures are absolute central-volume pressures. Nominal breakdown fields are calculated from \(E_{\mathrm{b}}=V_{\mathrm{b}}/d_{\mathrm{c}}\), using the cold electrode
gaps \(d_{\mathrm{c}}\) listed in Table~\ref{tab:runs}. For stationary datasets, entries labeled ``mean'' correspond to the full-distribution
means summarized in Table~\ref{tab:distribution-statistics}. The lower- and higher-field entries for the 602~Torr LHe dataset are empirical component means. Because the 612~Torr LAr dataset is non-stationary, its first-breakdown and approximate plateau values are listed instead of representing the dataset by a single stationary distribution mean.
}
\label{tab:principal-values}
\begin{tabular}{llcccc}
\toprule
Run & Liquid & $P$ [Torr] & $\Vb$ [kV] & $\Eb$ [kV cm$^{-1}$] & Note \\
\midrule
1 & \LNtwo & 600 & 25.4 & 751 & mean \\
1 & \LNtwo & 760 & 29.7 & 879 & mean \\
2 & \LHe   & 602 & 18.5 & 564 & all events, mean \\
2 & \LHe   & 602 & 12.2 & 372 & lower-field component mean \\
2 & \LHe   & 602 & 22.0 & 671 & higher-field component mean \\
2 & \LHe   & 768 & 25.0 & 762 & mean \\
3 & \LAr   & 612 & 7.72 & 226 & first breakdown \\
3 & \LAr   & 612 & 15.2 & 446 & approximate plateau \\
3 & \LAr   & 760 & 18.8 & 551 & mean \\
4 & \LNtwo & 602 & 21.4 & 633 & mean \\
4 & \LNtwo & 760 & 25.8 & 763 & mean \\
\bottomrule
\end{tabular}
\end{table}

Using the \SI{760}{Torr} data as an example, let
\(E_1=\SI{879}{kV.cm^{-1}}\) denote the mean breakdown field from
\LNtwo run~1 and \(E_4=\SI{763}{kV.cm^{-1}}\) denote the corresponding
mean field from \LNtwo run~4. The fractional run-to-run surface-state
variation is then estimated as
\begin{equation}
    \delta_{\mathrm{surf}}
    =
    \frac{|E_1-E_4|}
    {\frac{1}{2}(E_1+E_4)}
    =
    \frac{|879-763|}
    {(879+763)/2}
    \simeq 0.14 .
    \label{eq:surface-systematic}
\end{equation}
This 14\% variation is used below as a practical estimate of the systematic uncertainty associated with electrode-surface evolution. Equivalently, the run~4 field is lower than the run~1 field by
$(879 - 763)/879 \simeq 13\%$.  Note that the 14\%
value in Eq.~\eqref{eq:surface-systematic} is the symmetric fractional difference relative to the average of the two fields; the symmetric value is used hereafter.

In addition to the run-to-run shift, the LN$_2$ measurements from both cooldowns contain a small number of lower-field breakdown events. These events may indicate a distinct lower-field subset, but they are also consistent with the lower tail of a single continuous breakdown distribution. The present data do not clearly distinguish between these two interpretations. If an empirical component boundary is imposed, the lower-field events account for approximately 10\%--15\% of the sample, depending on cooldown, although this fraction is sensitive to the selected boundary. Therefore, we treat this structure only as a descriptive feature of the measured distributions and do not assign it to a separate physical population. Regardless of its origin, the lower-field tail should be considered when extrapolating breakdown probabilities to larger stressed areas, because system-level failure probability may be influenced more strongly by the lower tail of the distribution than by the mean breakdown field.


\subsection{Liquid helium}\label{sec:lhe-results}

\begin{figure}
    \centering
	\subfloat[LHe, 602 Torr]{\includegraphics[width=0.503\linewidth]{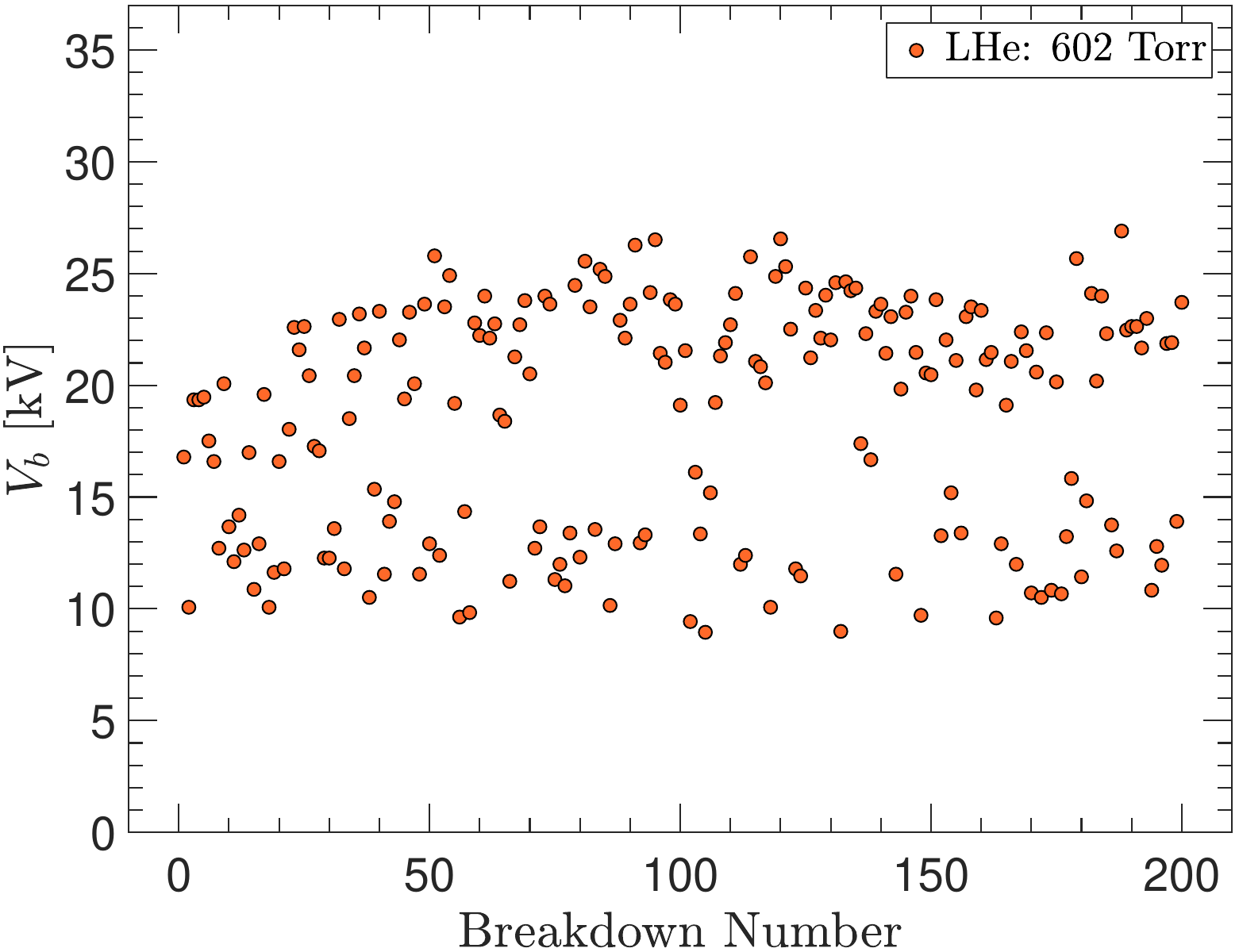}\label{fig:LHe-Vb-seq-602Torr}}
	\subfloat[LHe, 602 Torr]{\includegraphics[width=0.497\linewidth]{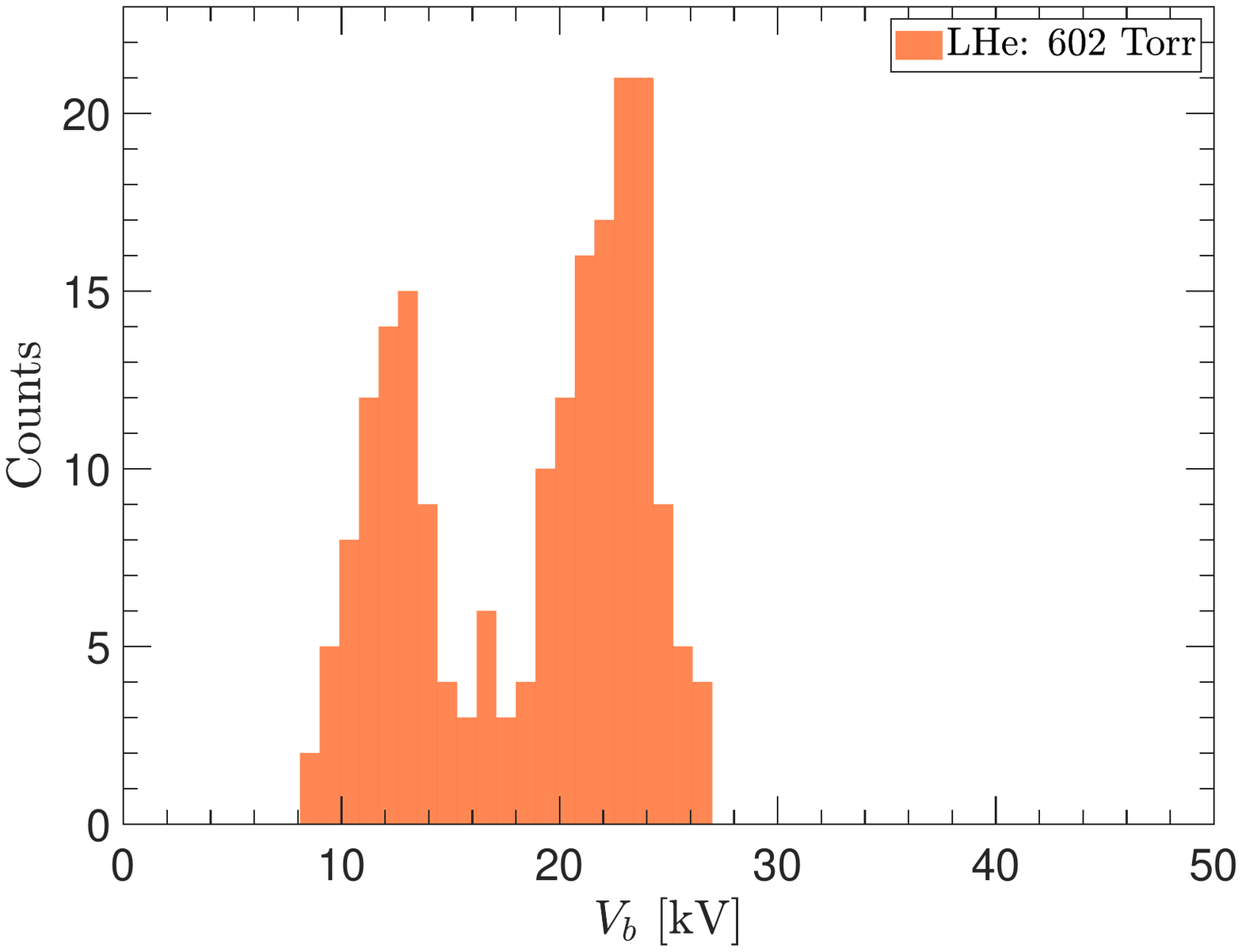}\label{fig:LHe-hist-602Torr}} \\
	\subfloat[LHe, 768 Torr]{\includegraphics[width=0.499\linewidth]{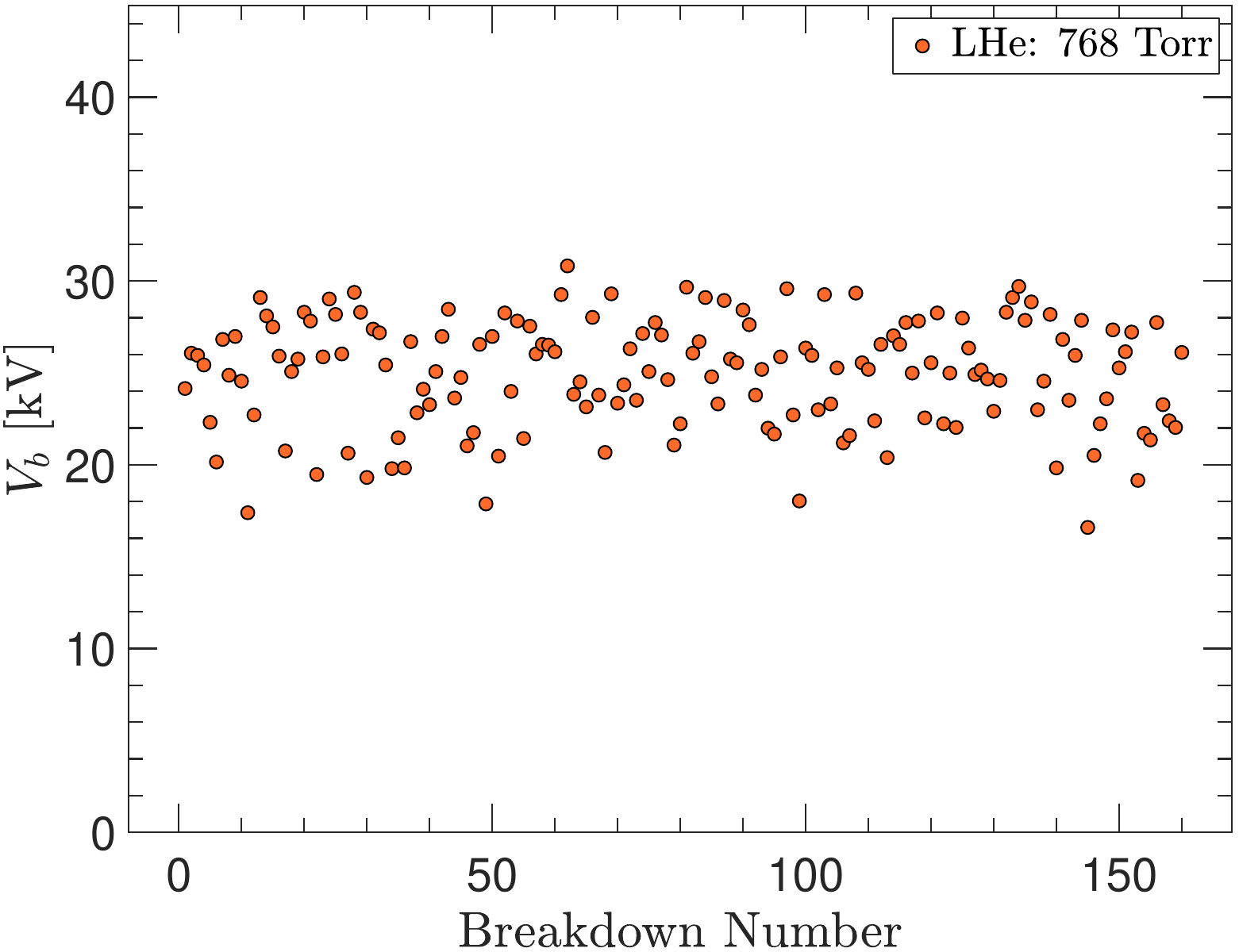}\label{fig:LHe-seq-768Torr}}
    \subfloat[LHe, 768 Torr]{\includegraphics[width=0.501\linewidth]{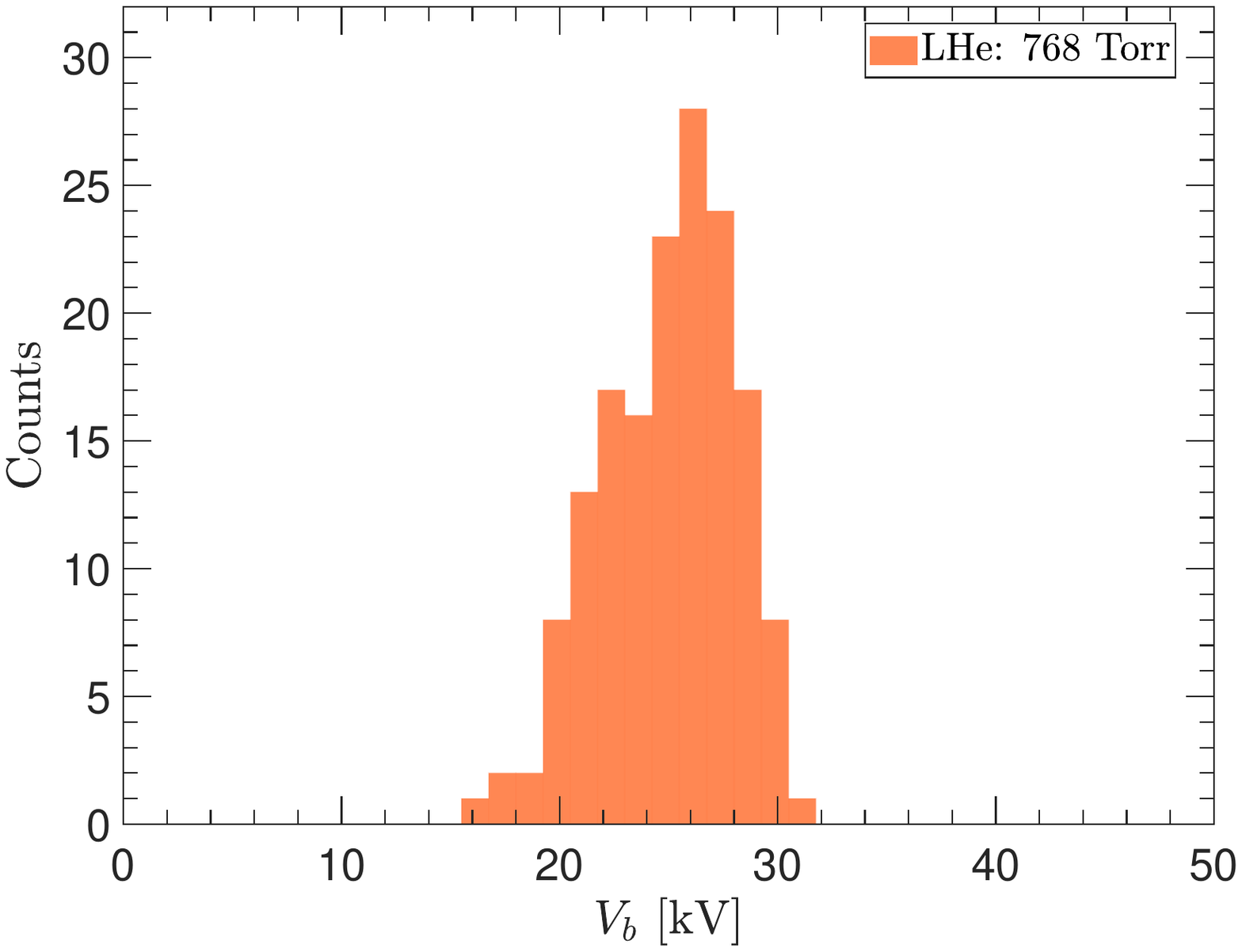}\label{fig:LHe-hist-768Torr}}
    \caption{Breakdown-voltage time series and histograms for \LHe run~2. Panels~(a,b) show the \SI{602}{Torr} dataset with $N=200$ breakdowns; panels~(c,d) show the \SI{768}{Torr} dataset with $N=160$ breakdowns. At \SI{602}{Torr}, the distribution is bimodal, with lower- and higher-field component means of approximately \SI{12.2}{kV} and \SI{22.0}{kV}. At \SI{768}{Torr}, the distribution shifts upward and the bimodality is less distinct, with a mean breakdown voltage of approximately \SI{25.0}{kV}.}
\label{fig:lhe}
\end{figure}

The dielectric behavior of \LHe, illustrated in Figure~\ref{fig:lhe}, shows a distinct bimodal distribution at \SI{602}{Torr} and \SI{3.99}{K}. The lower-field and higher-field components have mean breakdown voltages of approximately \SI{12.2}{kV} and \SI{22.0}{kV}, corresponding to fields of \SI{372}{kV.cm^{-1}} and \SI{671}{kV.cm^{-1}}, respectively. The ratio of the higher-field to lower-field component means is approximately
\begin{equation}
    \frac{\langle \Vb\rangle_{\mathrm{high}}}
         {\langle \Vb\rangle_{\mathrm{low}}}
    =
    \frac{22.0}{12.2}
    \simeq 1.80 .
    \label{eq:lhe-mode-ratio}
\end{equation}

\begin{table}[t]
\centering
\caption{
Empirical component summary for the \SI{602}{Torr} \LHe dataset. 
The component boundary at $\Vb=\SI{16.5}{kV}$ corresponds to the valley between the two histogram peaks in Figure~\ref{fig:lhe}. }
\label{tab:lhe-components}
\begin{tabular}{lcccc}
\toprule
Component & Selection & $N$ & $\langle\Vb\rangle$ [kV] & $\langle\Eb\rangle$ [kV cm$^{-1}$] \\
\midrule
Lower-field & $\Vb \leq \SI{16.5}{kV}$ & 72 & 12.2 & 372 \\
Higher-field & $\Vb > \SI{16.5}{kV}$ & 128 & 22.0 & 671 \\
\bottomrule
\end{tabular}
\end{table}

The component definitions and corresponding mean values are summarized in Table~\ref{tab:lhe-components}. At 768~Torr, the voltage histogram no longer exhibits the same clear bimodal separation. The mean breakdown voltage increases to approximately 25.0~kV, corresponding to $762~\mathrm{kV\,cm^{-1}}$ using the cold gap in Table~\ref{tab:runs}. Pressurization shifts the LHe distribution upward and makes the bimodality less distinct. However, the reduced bimodality at 768~Torr does not imply that lower-field events are eliminated. A more detailed discussion of this behavior is given in Section~\ref{sec:lhe-bimodal}.

\begin{figure}
    \centering
	\subfloat[LAr, 612 Torr]{\includegraphics[width=0.503\linewidth]{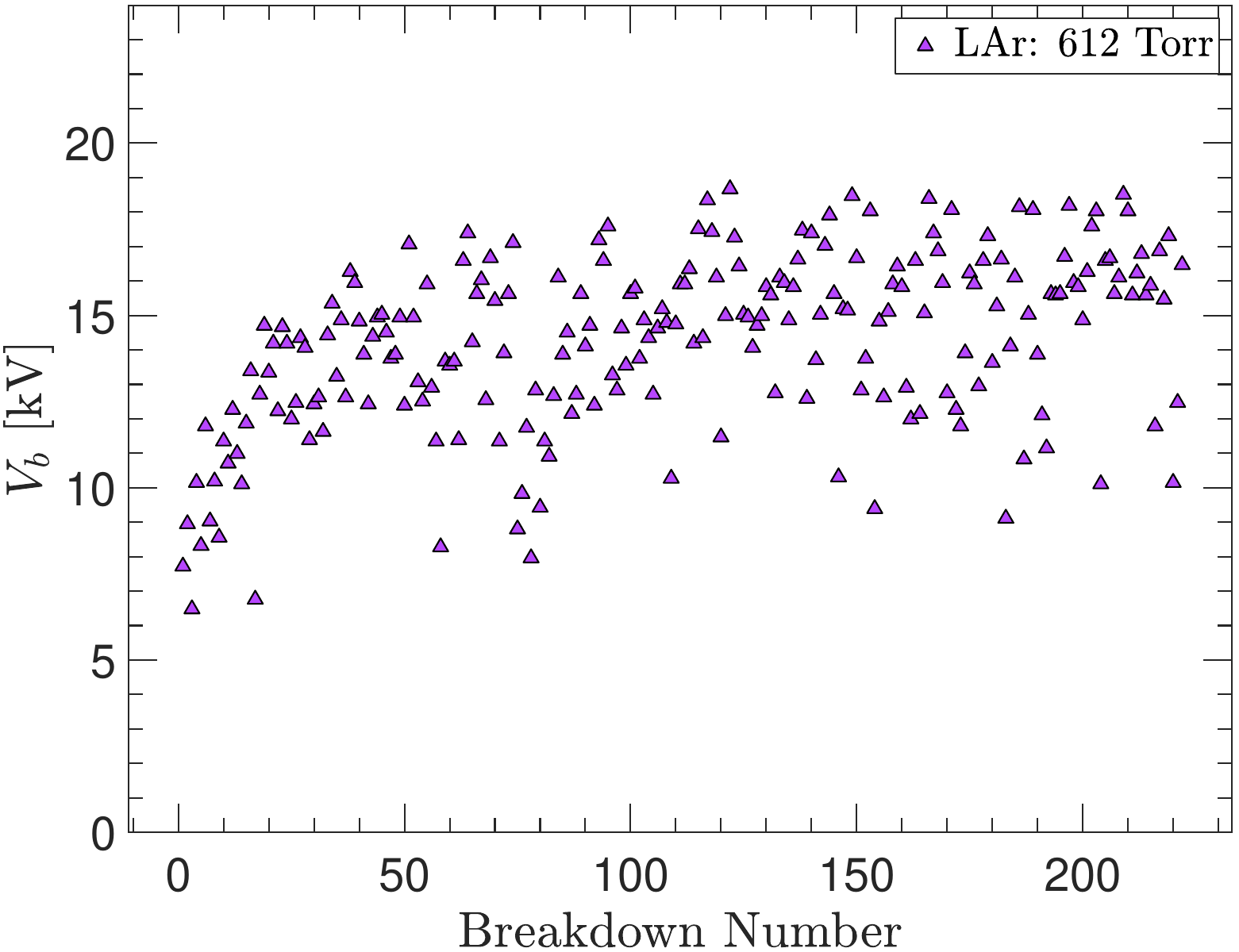}\label{fig:LAr-Run3-seq-612Torr}}
	\subfloat[LAr, 612 Torr]{\includegraphics[width=0.497\linewidth]{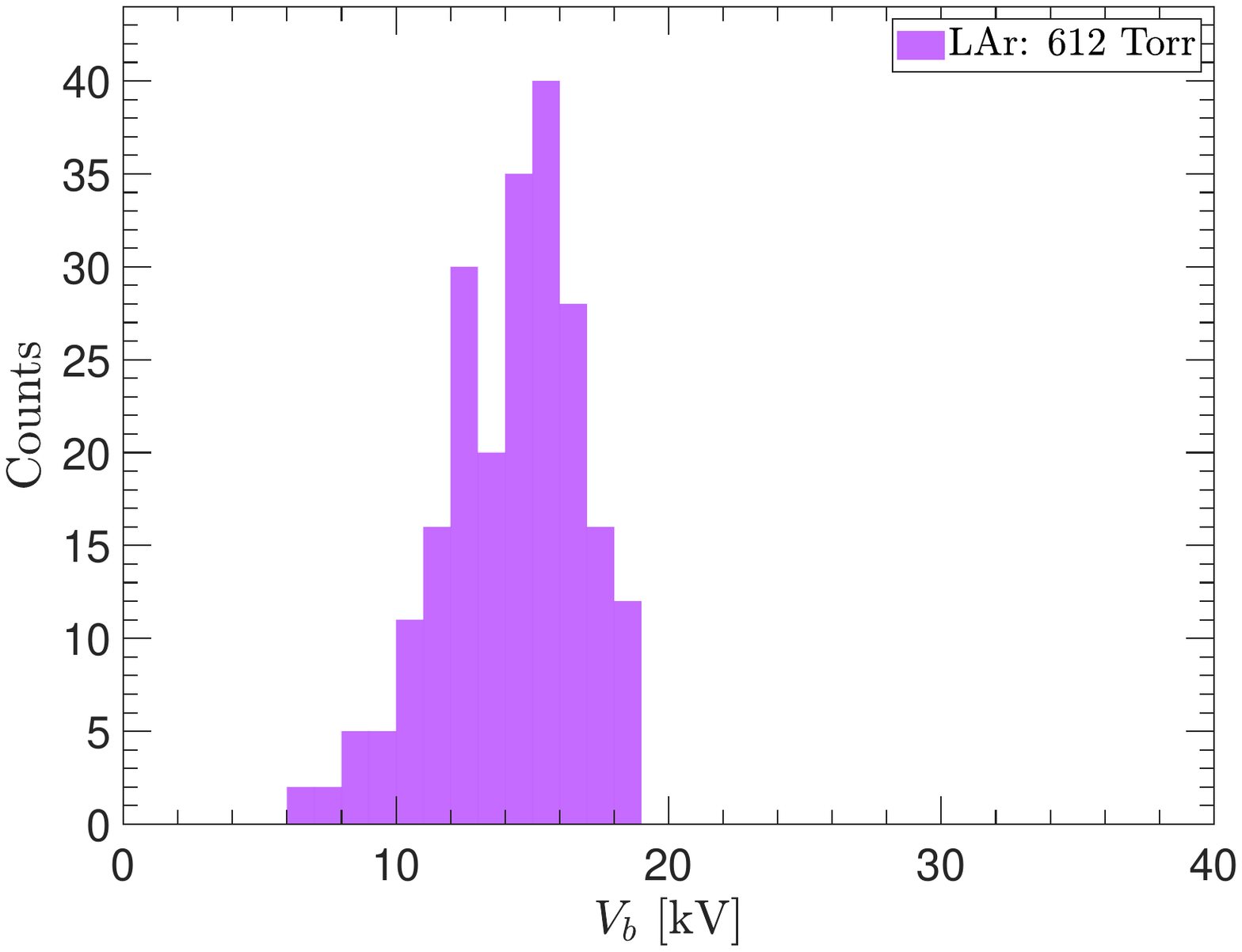}\label{fig:LAr-Run3-hist-612Torr}} \\
	\subfloat[LAr, 760 Torr]{\includegraphics[width=0.503\linewidth]{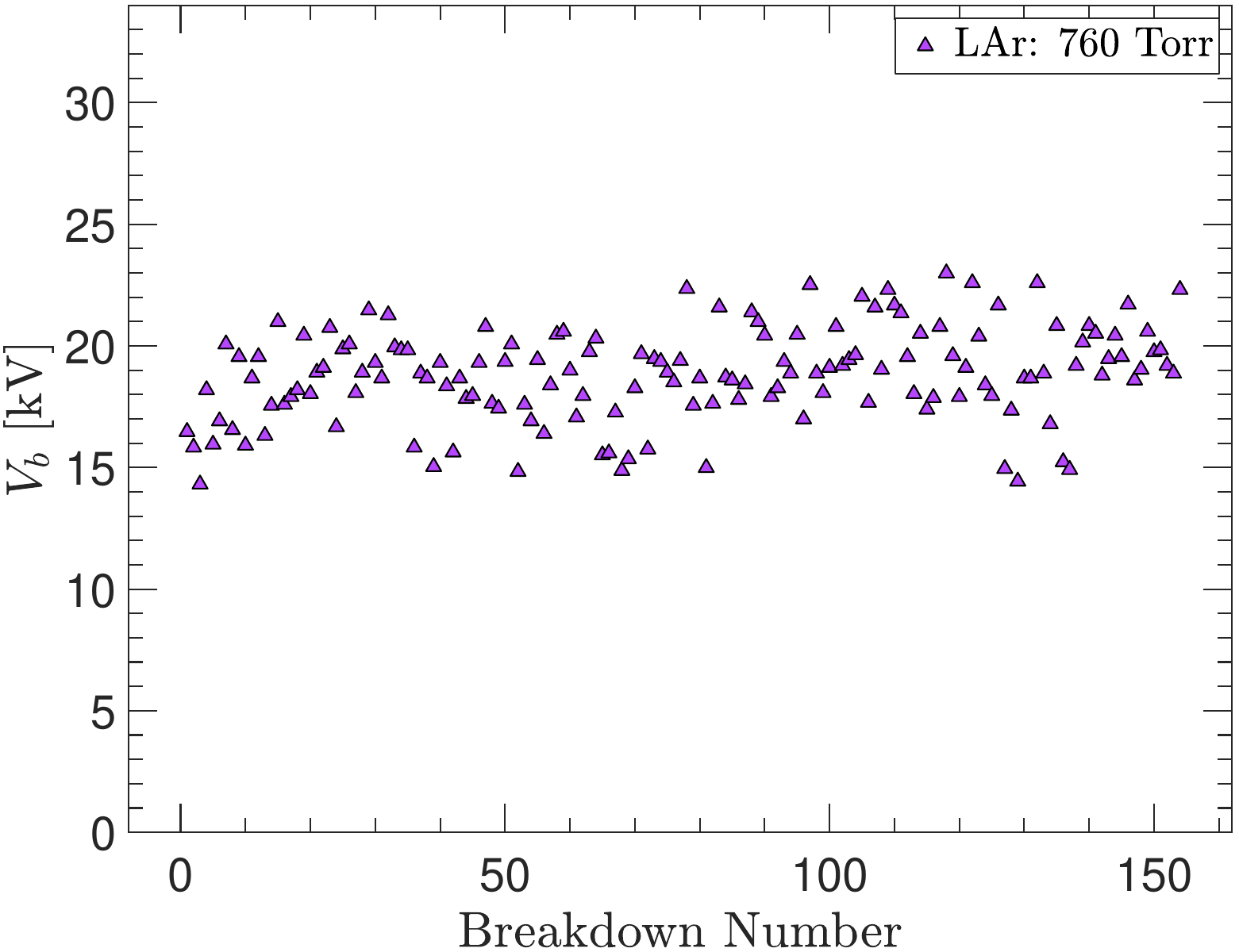}\label{fig:LAr-Run3-seq-760Torr}}
    \subfloat[LAr, 760 Torr]{\includegraphics[width=0.497\linewidth]{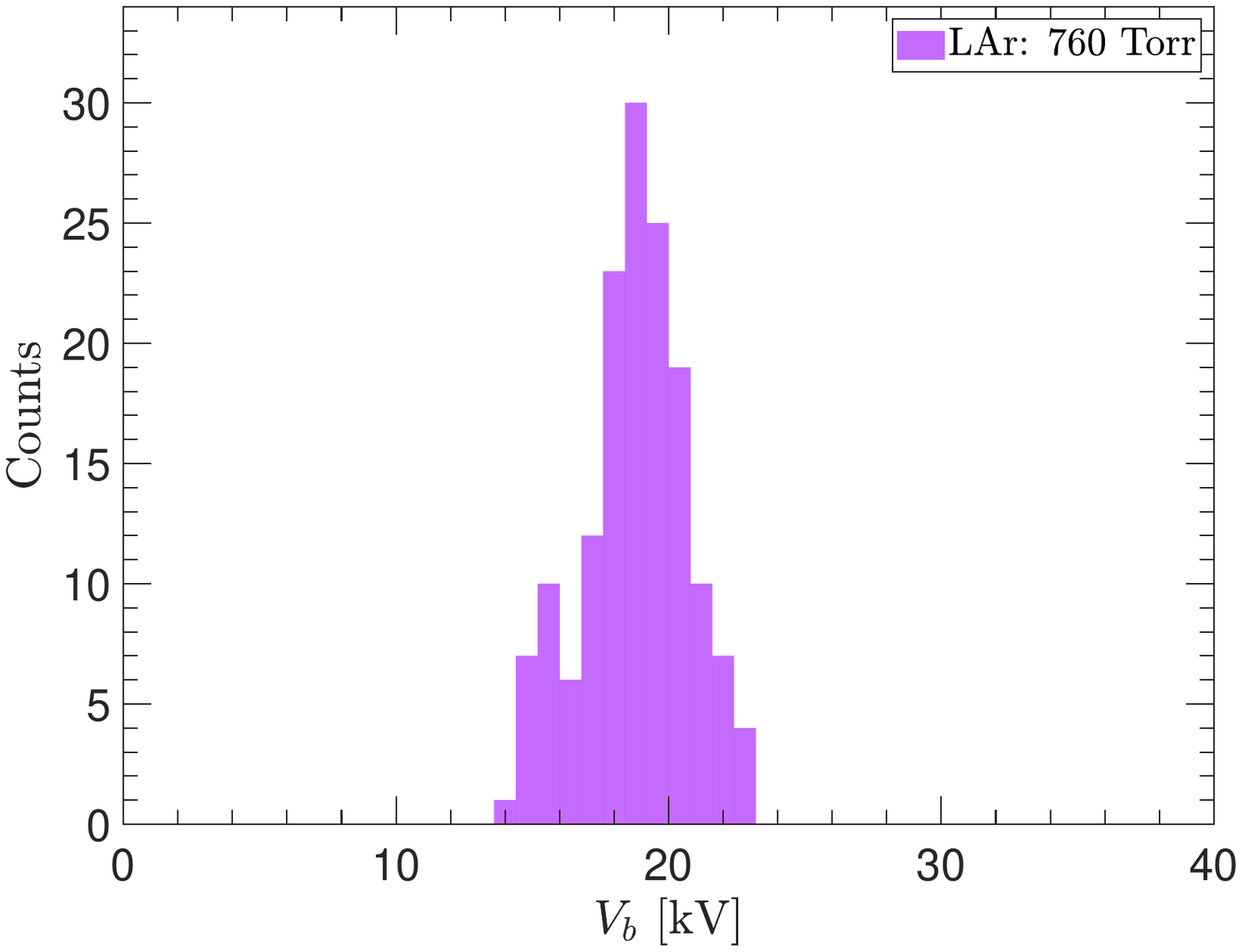}\label{fig:LAr-Run3-hist-760Torr}}
    \caption{Breakdown-voltage time series and histograms for \LAr run~3. Panels~(a,b) show the \SI{612}{Torr} dataset with $N=222$ breakdowns; panels~(c,d) show the \SI{760}{Torr} dataset with $N=154$ breakdowns. The \SI{612}{Torr} dataset is non-stationary and shows a turn-on feature: the first breakdown occurred at \SI{7.72}{kV}, while the approximate plateau voltage is \SI{15.2}{kV}. The \SI{760}{Torr} dataset is narrower and has a mean breakdown voltage of approximately \SI{18.8}{kV}.}
\label{fig:lar}
\end{figure}


\subsection{Liquid argon}
\label{sec:lar-results}

The dielectric response of \LAr at \SI{612}{Torr}, shown in Figure~\ref{fig:lar}, exhibits an initial increase in breakdown voltage before reaching an approximately stationary plateau. We refer to this behavior as a turn-on feature. The plateau value quoted below is estimated from the approximately stationary portion of the time series after the initial rise rather than from a fit to a time-dependent model. Similar behavior has been reported in prior \LAr breakdown studies~\cite{Lockwitz2019,Blatter2014}. In the present apparatus, one possible explanation is that repeated breakdowns modify the local impurity environment or electrode surface state in the high-field region. Electronegative species could be redistributed or released from electrode oxides, adsorbed surface layers, or other local contamination and then remain temporarily confined within the CV. Such impurity evolution would reduce the effective mobility of free electrons by converting them into less mobile negative ions, while surface-state changes
could also modify local charge injection. Either effect could alter the measured breakdown voltage; in the observed direction, reduced carrier mobility or reduced charge injection would tend to increase it.

This hypothesis could not be directly tested because the present measurement did not include an in situ impurity monitor at the electrode gap or an in situ surface
diagnostic. The impurity concentration in the bulk supply gas or liquid may differ from the local concentration in the high-field region, and both the local impurity environment and electrode surface state may evolve after each
discharge. Therefore, the impurity- and surface-evolution explanation is treated here only as a possible mechanism.

The first breakdown in the \SI{612}{Torr} \LAr dataset occurred at $\Vb=\SI{7.72}{kV}$, while the approximate plateau voltage was $\Vb\simeq\SI{15.2}{kV}$, giving
\begin{equation}
    \frac{\Vb^{\mathrm{plateau}}}{\Vb^{\mathrm{first}}}
    =
    \frac{15.2}{7.72}
    \simeq 1.97 .
    \label{eq:lar-turnon-ratio}
\end{equation}
The definition of the turn-on and plateau regions is summarized in Table~\ref{tab:lar-turnon}. This ratio is similar to that reported by Lockwitz and Jostlein~\cite{Lockwitz2019}, although the comparison is qualitative because the impurity history, electrode geometry, and liquid-exchange conditions differ between the experiments.

\begin{table}[t]
\centering
\caption{
Operational definition of the turn-on and plateau regions in the \SI{612}{Torr} \LAr dataset. 
The plateau voltage quoted in Table~\ref{tab:principal-values} is calculated from the plateau subset defined here.
}
\label{tab:lar-turnon}
\begin{tabular}{lcccc}
\toprule
Region & Event range & $N$ & $\langle\Vb\rangle$ [kV] & $\langle\Eb\rangle$ [kV cm$^{-1}$] \\
\midrule
Turn-on & 1–50 & 50 & 12.5 & 367 \\
Plateau & 173–222 & 50 & 15.2 & 446 \\
\bottomrule
\end{tabular}
\end{table}

Approximately two hours elapsed between the end of the 612~Torr sequence and the start of the 760~Torr sequence due to the time required to pressurize the CV. Therefore, the 760~Torr data are not a temporally continuous extension of the 612~Torr time series. During this interval, the local impurity distribution could have evolved, whereas more persistent modifications of the electrode surface could have remained.

At 760~Torr, the pronounced initial turn-on observed at 612~Torr is not present, and the distribution is narrower, with a mean breakdown voltage of approximately \(18.8~\mathrm{kV}\), corresponding to \(551~\mathrm{kV\,cm^{-1}}\). The time series nevertheless exhibits a weak positive drift with breakdown number. This gradual increase is small relative to the event-to-event spread and is therefore better described as a residual drift than as a distinct second turn-on. It is qualitatively consistent with the continued evolution of the local impurity environment or electrode surface state after pressurization. However, if the surface-state evolution acted in the direction observed by Swan and Lewis~\cite{Swan1960} — where reduction of the oxide layer lowered the breakdown field — it would imply a
negative slope, contrary to the weak positive drift seen here. The \SI{760}{Torr} distribution also contains a lower-voltage tail, although the present analysis does not assign this tail to a separate physical component.


\section{Discussion}\label{sec:discussion}

\subsection{Comparison of the liquids}
\label{sec:liquid-comparison}

The principal breakdown voltages and corresponding nominal breakdown fields used in the cross-liquid comparison are summarized in Table~\ref{tab:principal-values}. At approximately 760~Torr, the mean gap-corrected breakdown fields measured in this work are
\begin{equation}
\begin{aligned}
E_{\mathrm{b}}(\mathrm{LN_2,\ run\ 1}) &\simeq 879~\mathrm{kV\,cm^{-1}},\\
E_{\mathrm{b}}(\mathrm{LN_2,\ run\ 4}) &\simeq 763~\mathrm{kV\,cm^{-1}},\\
E_{\mathrm{b}}(\mathrm{LHe})            &\simeq 762~\mathrm{kV\,cm^{-1}},\\
E_{\mathrm{b}}(\mathrm{LAr})            &\simeq 551~\mathrm{kV\,cm^{-1}}.
\end{aligned}
\label{eq:field-comparison}
\end{equation}
The values in Eq.~\eqref{eq:field-comparison} correspond to the approximately 760~Torr entries in
Table~\ref{tab:principal-values}. They are calculated from the measured mean breakdown voltages and the cold electrode gaps listed in Table~\ref{tab:runs}. The dominant instrumental contribution to the absolute field scale uncertainty is the cold-gap uncertainty of approximately 4\%. This primarily limits the absolute calibration of the reported breakdown fields but not the relative comparison between liquids in the present study. In contrast, the empirical run-to-run surface-state variation, \(\delta_{\mathrm{surf}}\simeq14\%\), limits how strongly differences between liquids can be interpreted as dielectric material-dependent rather than surface-history-dependent.

With these uncertainties in mind, pressurized \LHe and \LNtwo have comparable breakdown fields in the present apparatus. The \SI{768}{Torr} \LHe mean field is nearly identical to the \SI{760}{Torr} field from \LNtwo run~4 and lies within the observed surface-state variation (14\%) of the mean breakdown field measured for \LNtwo run~1. Therefore, the difference between pressurized LHe and LN$_2$ is not resolvable within the observed surface-state variation and field-scale uncertainty.

The comparison with \LAr is more robust. Using the two \LNtwo cooldowns as an empirical bracket, the \SI{760}{Torr} \LNtwo field exceeds the \LAr field by approximately
\begin{equation}
    \frac{763-551}{551} \simeq 39\%
    \quad\mathrm{to}\quad
    \frac{879-551}{551} \simeq 60\%.
\end{equation}
Similarly, the \SI{768}{Torr} \LHe field exceeds the \SI{760}{Torr} \LAr field by
\begin{equation}
    \frac{762-551}{551}
    \simeq 38\%.
\end{equation}
These differences are larger than the observed 14\% surface-state variation and remain substantial even after allowing for the absolute field-scale
uncertainty. They indicate that, under the common geometry and voltage-ramp procedure used here, \LAr breaks down at a lower field than either \LNtwo or pressurized \LHe.

The comparison above addresses the pressurized datasets near
\(\SI{760}{Torr}\). The interpretation changes when the lower-pressure \LHe dataset is included, because that dataset contains a strong lower-field component. The comparison between \LNtwo and \LHe depends strongly on whether \LHe is measured near the saturated-vapor-pressure condition or under pressurization. The lower-field LHe component at 602~Torr lies well below the corresponding
lower-pressure LN$_2$ distributions from runs~1 and~4, whereas the 768~Torr LHe mean field is comparable to the 760~Torr LN$_2$ fields within the field-scale and surface-state uncertainties. The contrast with \LAr is larger and remains the clearest cross-liquid separation observed in this apparatus.

A limitation of the comparison is that the liquid identity is correlated with run order. The same electrode pair was intentionally used throughout the experiment to remove geometry and surface-finish differences between liquids, but this also means that the electrode surface evolved through cumulative breakdown exposure. The repeat \LNtwo measurement in run~4 provides a direct estimate of this effect. Therefore, cross-liquid differences comparable to the empirical
14\% surface-state variation should be interpreted with caution. Differences much larger than this uncertainty are less likely to be explained solely by electrode-surface evolution or by the nominal field calibration uncertainty.


\subsection{Negative-carrier transport and dielectric strength}
\label{sec:mobility}

Experimental measurements suggest an empirical relationship between dielectric breakdown strength and the transport state of the dominant negative charge carrier. The clearest distinction is between liquids in which injected electrons remain quasi-free, such as LAr, LKr, and LXe, and liquids in which the negative charge carriers are localized as electron bubbles or molecular ions, such as \LHe, LNe, LH$_2$, \LNtwo, and LO$_2$. The quasi-free-electron liquids have high mobilities in the low-field regime and generally exhibit lower measured breakdown fields, whereas liquids with
localized negative carriers have much lower mobilities and generally exhibit higher breakdown fields.

For breakdown to occur, an initiating seed, such as field-emitted charge, must evolve into a self-sustaining process before the carriers are trapped or
neutralized. If mobility is treated as a proxy for the ability of injected charge to feed a prebreakdown process, then, for comparable injected-carrier populations and applied fields, a larger drift response produces a larger prebreakdown current and more rapid charge transport. This can facilitate charge multiplication, local field modification, or field-emission-assisted heating. In contrast, localization, strong scattering, and attachment impede these processes and therefore tend to require a larger macroscopic field before a self-sustaining instability can develop.

These considerations motivate a phenomenological inverse relationship between breakdown field, $E_{\mathrm{b}}$, and an effective mobility, $\mu_{\mathrm{eff}}$, that is characteristic of the prebreakdown charge carrier population. Hence, this trend may be represented as
\begin{equation}
    E_{\mathrm{b}}
    \propto
    \mu_{\mathrm{eff}}^{-\alpha},\ \ \alpha >0,
    \label{eq:mobility-scaling}
\end{equation}
where $\mu_{\mathrm{eff}}$ denotes an effective transport response under the high-field conditions preceding breakdown, and $\alpha$ is an empirical exponent. 

Impurity studies provide independent support for the expected relationship between charge-carrier mobility and breakdown strength. In high-mobility media such as LAr and LXe, electronegative impurities capture quasi-free electrons and convert the dominant charge carriers into much less mobile negative ions. This process can reduce the effective negative-charge mobility by up to six
orders of magnitude; for example, the mobility of $\mathrm{O_2^-}$ in LXe is
approximately $7\times10^{-4}\,\mathrm{cm^2\,V^{-1}\,s^{-1}}$~\cite{Hilt1994}. Consistent with this reduction in mobility, increasing oxygen concentration has been associated with a two- to three-fold increase in breakdown voltage~\cite{Swan1961}. The turn-on behavior observed in the 612 Torr LAr data may reflect a related mechanism: repeated discharges could modify the local impurity environment or electrode surface in the high-field region, thereby changing the effective charge-transport conditions and, consequently, the breakdown behavior. By contrast, electronegative impurities have a weaker effect in LN$_2$ and LHe, where negative-charge transport is already characterized by intrinsically low mobilities. Therefore, the smaller relative change in carrier mobility is consistent with the observed insensitivity of their breakdown strengths to impurity concentration~\cite{Yoshino1982}.

 \begin{table}[htb]
\centering
\caption{
Low-field mobilities of the dominant negative charge carriers in selected cryogenic liquids. These values are used only as comparative indicators of the carrier-transport regime and are not assumed to
represent the nonlinear transport response under breakdown conditions.
}
\label{tab:mobility}
\begin{tabular}{lccc}
\toprule
Cryogen & $T$ [K] & Dominant negative carrier state & $\mu_-$ [\si{cm^2.V^{-1}.s^{-1}}] \\
\midrule
Helium~\cite{Meyer1962}     & 4.2 & localized electron bubble & $1.98$--$2.35\times10^{-2}$ \\
Neon~\cite{Sakai1992}       & 25  & localized electron bubble & $1.4\times10^{-3}$ \\
Argon~\cite{Miller1968}     & 85  & quasi-free electron & $475$--$625$ \\
Krypton~\cite{Miller1968}   & 117 & quasi-free electron & $1800$ \\
Xenon~\cite{Miller1968}     & 163 & quasi-free electron & $2200$ \\
Nitrogen~\cite{Gee1985}     & 77  & localized negative ion & $2.5\times10^{-3}$ \\
Oxygen~\cite{Loveland1972}  & 90  & localized negative ion & $1.5\times10^{-3}$ \\
Hydrogen~\cite{Levchenko1992} & 21 & localized electron bubble & $1.5\times10^{-2}$ \\
\bottomrule
\end{tabular}
\end{table}

Table~\ref{tab:mobility} lists the low-field mobility values for a range of common cryogenic liquids. It is important to note that low-field mobility characterizes near-equilibrium charge transport, whereas breakdown develops in a strongly non-equilibrium, high-field regime. Therefore, it is more appropriate to treat the tabulated mobility as a proxy for carrier localization and transport regime, rather than as the microscopic quantity that directly sets the breakdown threshold. As such, the low-field mobility distinguishes liquids containing quasi-free electrons from those in which the negative charge carriers are localized or strongly scattered. Notably, variations in the measured breakdown fields between liquids are much smaller than the several-order-of-magnitude range in the tabulated low-field mobilities.

\begin{figure}[htb]
    \centering
    {\includegraphics[width=0.6\linewidth]{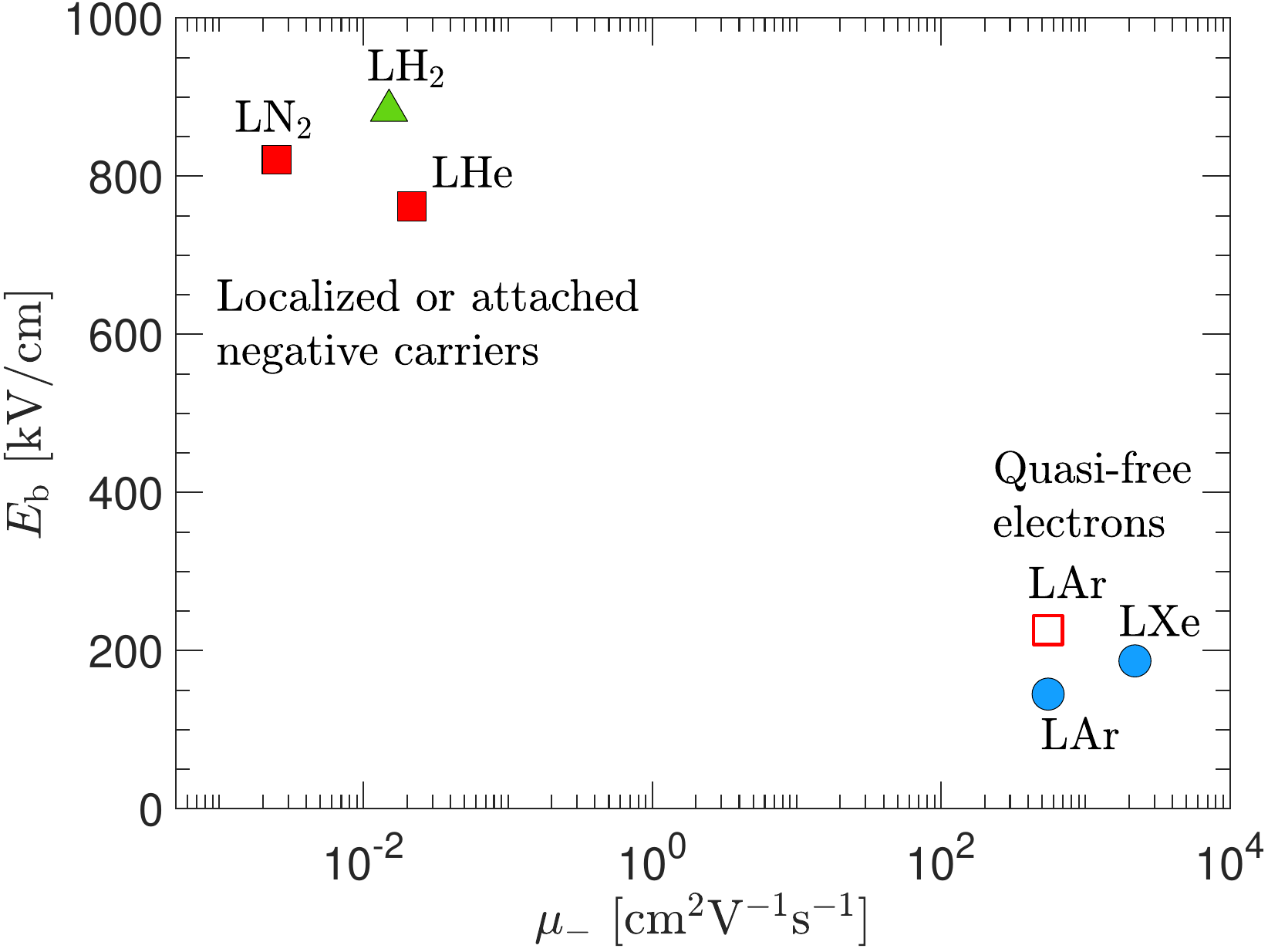}}\label{fig:Eb_vs_mobility}
\caption{DC electrical breakdown field as a function of the low-field mobility of the dominant negative charge carrier for selected cryogenic liquids. Square markers show measurements from this work: the LN$_2$ point is the average of the 760~Torr mean fields from runs~1 and~4, the LHe point is the mean of the 768~Torr dataset, and the LAr point is the first breakdown at 612~Torr plotted at the nominal pure-LAr mobility. Circular markers show literature measurements for LAr and LXe~\cite{Tvrznikova2019} obtained at stressed areas and pressures comparable to those of the present work. The triangular LH$_2$ point is an indirect estimate obtained by scaling the relative LN$_2$--LH$_2$ comparison of Jefferies and Mathes~\cite{Jefferies1970} to the LN$_2$ breakdown field measured in this work.}
\label{fig:mobility}
\end{figure}

Figure~\ref{fig:mobility} compares the measured breakdown fields with the low-field negative-carrier mobilities summarized in Table~\ref{tab:mobility}, together with representative breakdown measurements from the literature. The present LAr first-breakdown measurement at 612 Torr is shown as an open symbol and provides a pre-turn-on reference obtained in the same apparatus as the LN$_2$ and LHe measurements. The point is plotted at the nominal pure-LAr low-field mobility for reference. Because the effective carrier transport was not independently measured and may have evolved with impurity or surface conditions, its horizontal position should not be interpreted as a measured mobility. The LAr point from Ref.~\cite{Tvrznikova2019} is used instead to represent the high-mobility regime at a comparable stressed area. Taken together, the data are consistent with a qualitative inverse trend between breakdown field and mobility and suggest two broad carrier-transport regimes. Liquids with high-mobility quasi-free electrons occupy the lower-breakdown-field region, whereas liquids with localized or strongly scattered negative carriers generally occupy the higher-field region. The scatter within these groups, together with the strong pressure dependence observed in LHe, indicates that negative-carrier transport is only one part of the breakdown mechanism. Other factors such as thermodynamic stability against bubble formation, impurity content, and electrode surface state are similarly important.

This phenomenological trend has practical consequences. For HV insulation, localization or attachment of negative charge may be beneficial because it limits rapid energy accumulation and charge multiplication. For noble-liquid detectors, the same high mobility that enables efficient charge collection may impose more stringent constraints on HV design. Thus, the comparison highlights a fundamental tension between charge-transport performance and dielectric robustness in cryogenic liquids.


\subsection{Bubble-mediated breakdown and bimodality in LHe}
\label{sec:lhe-bimodal}

As shown in Figure~\ref{fig:lhe}, the LHe dataset measured at
\(602~\mathrm{Torr}\) and \(3.99~\mathrm{K}\) exhibits a pronounced bimodal breakdown-voltage distribution. The lower- and higher-field components have mean breakdown voltages of approximately \(12.2~\mathrm{kV}\) and
\(22.0~\mathrm{kV}\), corresponding to fields of
\(372~\mathrm{kV\,cm^{-1}}\) and \(671~\mathrm{kV\,cm^{-1}}\), respectively.  At \(768~\mathrm{Torr}\), the distribution shifts to higher field and the bimodality is much less distinct. This pressure response is the strongest evidence in this dataset that bubble nucleation or bubble growth contributes to the lower-field component.

To compare the two components without assuming a particular parametric distribution, we adopt the empirical cumulative-hazard representation of Phan~\textit{et al.}~\cite{Phan2021}, derived from general statistical considerations of breakdown phenomena. Let \(\widehat{P}_b(\Eb)\) be the empirical cumulative probability that breakdown has occurred by field \(\Eb\), where the hat denotes an empirical estimate. The empirical survival probability is then
\begin{equation}
    \widehat{P}_s(\Eb)
    =
    1-\widehat{P}_b(\Eb),
    \label{eq:empirical-survival}
\end{equation}
and the empirical cumulative hazard function is
\begin{equation}
    \widehat{H}(\Eb)
    =
    -\ln \widehat{P}_s(\Eb)
    =
    -\ln\left[1-\widehat{P}_b(\Eb)\right].
    \label{eq:cumulative-hazard}
\end{equation}

\begin{figure}[htb]
    \centering
	\subfloat[Cumulative hazard function]{\includegraphics[width=0.49\linewidth]{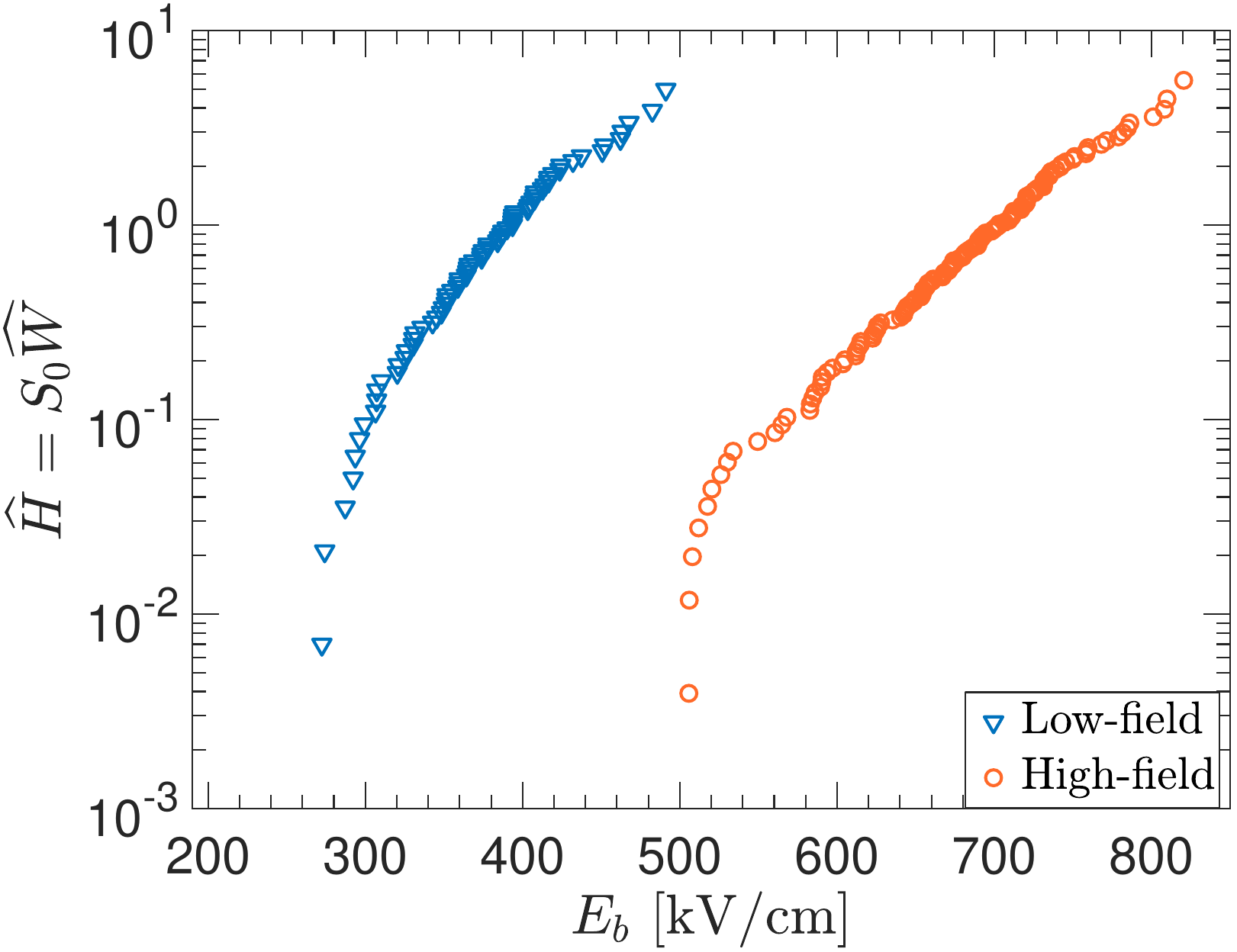}\label{fig:SW_vs_E_compare_LHe_600Torr}} \hspace{0.2cm}
    \subfloat[Fowler-Nordheim plot]{\includegraphics[width=0.49\linewidth]{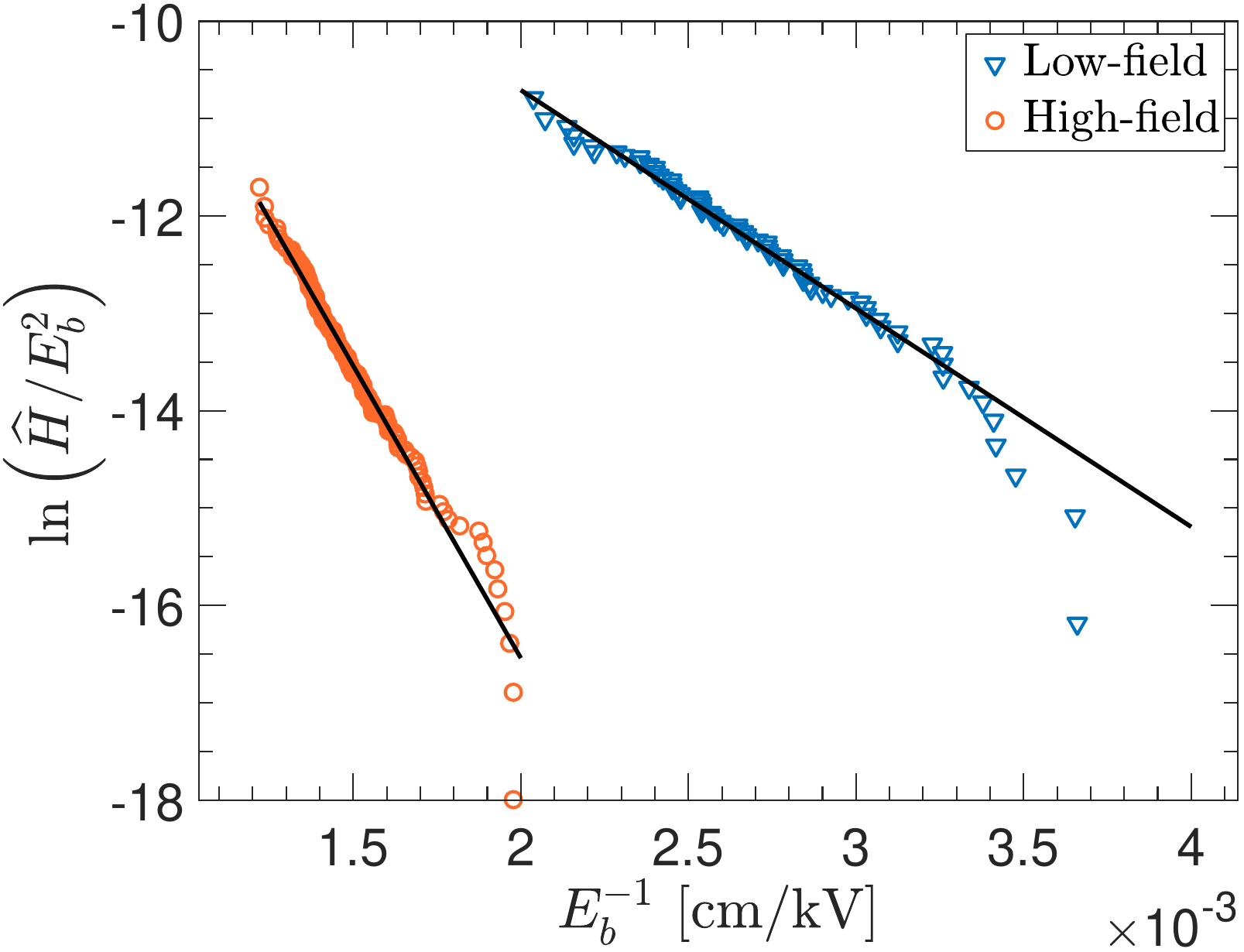}\label{fig:FN-plot-LHe-600Torr}}
    \caption{Empirical cumulative-hazard comparison for the lower- and higher-field
components of the \SI{602}{Torr} \LHe dataset. Panel~(a) shows the empirical
cumulative hazard \(\widehat{H}(\Eb)=S_0\widehat{W}(\Eb)\) as a function of
breakdown field. Panel~(b) shows the same data in a Fowler--Nordheim
representation, treating \(\widehat{H}\) as proportional to an effective
field-dependent initiation current. The component boundary is $V_b = 16.5$~kV, corresponding to the valley between the two peaks in Figure~\ref{fig:lhe}. The approximately parallel behavior indicates similar empirical field
dependence over the measured range.}
\label{fig:lhe-hazard}
\label{sec:lhe-results-bimodal}
\end{figure}

Because breakdown initiation is expected to occur predominantly at favorable electrode-surface sites, rather than through homogeneous nucleation distributed throughout the bulk liquid, a surface-based weakest-link description is appropriate. Unlike the empirical quantities above, the model survival probability, $P_s$, is written without a hat. The survival probability for the electrode surface is
\begin{equation}
    P_s(\Eb)
    =
    \exp\left[
    -\int_S W(E(\mathbf{r}))\,dS
    \right].
    \label{eq:survival-general}
\end{equation}
Here, $W$ is the local cumulative-hazard density per unit electrode area, and $\mathbf{r}$ denotes position on the electrode surface $S$. Physically, $W$ represents the local contribution to the breakdown hazard from surface-associated weak sites, including asperities, contaminants, microscopic cavities, and regions of enhanced field-emission heating. For an approximately uniform stressed area \(S_0\), this reduces to
\begin{equation}
    \widehat{H}(\Eb)
    \simeq
    S_0\widehat{W}(\Eb).
    \label{eq:hazard-area}
\end{equation}
Thus, $\widehat{H}$ is the empirical cumulative hazard, while
$\widehat{W}$ is the corresponding empirical cumulative-hazard density per unit stressed area. In Figure~\ref{fig:lhe-hazard}, we plot \(\widehat{H} = S_0\,\widehat{W}\) to compare the field dependence of the lower- and higher-field components.

Field emission from microscopic cathode asperities provides a plausible initiation mechanism for electrical breakdown in LHe. In this picture, the local field at an asperity is enhanced by a factor $\beta$, and the Fowler--Nordheim current may be written as
\begin{equation}
    I_{\mathrm{FN}}
    =
    A_e
    \frac{1.54}{\phi}
    10^{4.52\phi^{-1/2}}
    (\beta E)^2
    \exp\left[
    -\frac{6.53\times10^4\phi^{3/2}}{\beta E}
    \right],
    \label{eq:fowler-nordheim}
\end{equation}
where \(A_e\) is the effective emission area in \(\mathrm{cm^2}\), \(E\) is the applied macroscopic field in \(\mathrm{kV\,cm^{-1}}\), \(\phi\) is the electrode work function in eV, and \(\beta\) is the local field-enhancement factor~\cite{Wang1997}. Following Phan \textit{et al.}~\cite{Phan2021}, we use the Fowler--Nordheim field dependence as a phenomenological model for the cumulative hazard by taking
$\widehat{H}(E_b) \propto I_{\mathrm{FN}}$.

Under this assumption, the lower- and higher-field components are approximately linear in Fowler--Nordheim coordinates (panel~(b) of Figure~\ref{fig:lhe-hazard}), consistent with a common field-emission-related contribution to their initiation. Multiplying the field axis of the lower-field component in panel~(a) of Figure~\ref{fig:lhe-hazard} by the ratio of the component means, approximately 1.8, brings the two cumulative-hazard curves into approximate alignment. This behavior suggests that the two populations are not governed by unrelated mechanisms, but instead reflect a similar field-dependent initiation process
operating under different local thermodynamic or surface conditions. A similar field ratio of approximately 2.0 separates the pressurized
($1.7$~K, 612~Torr, $E_b = 551$~kV\,cm$^{-1}$) and low-pressure
($1.7$~K, 10~Torr, $E_b = 271$~kV\,cm$^{-1}$) LHe measurements of
Phan \textit{et al.}~\cite{Phan2021}.

The observed pressure dependence provides evidence that vapor formation contributes to the initiation of breakdown in LHe. To illustrate how pressure affects the nucleation barrier, it is useful first to consider homogeneous nucleation in an ideal, defect-free bulk liquid. For a spherical vapor bubble within a locally superheated liquid, the homogeneous nucleation barrier is
\begin{equation}
    \Delta G_{\mathrm{hom}}
    =
    \frac{16\pi\sigma^3}
    {3\left(\Delta P_{\mathrm{drive}}\right)^2},
    \label{eq:homogeneous-barrier}
\end{equation}
where \(\sigma\) is the liquid--vapor surface tension and
\(\Delta P_{\mathrm{drive}}\) is the effective pressure driving bubble growth~\cite{Fisher1949}. In this idealized picture, homogeneous nucleation incurs the full liquid--vapor interfacial free-energy cost and therefore provides a reference
barrier against which heterogeneous nucleation can be compared.

However, under practical HV conditions, vapor formation is more likely to initiate
heterogeneously at the electrode--liquid interface, where pre-existing surface
features can reduce the free-energy cost of forming a critical vapor nucleus.
In the classical spherical-cap approximation for nucleation on a locally flat
surface~\cite{Blander1975}, the heterogeneous nucleation barrier is
\begin{equation}
    \Delta G_{\mathrm{het}}
    =
    f(\theta)\Delta G_{\mathrm{hom}},
    \qquad
    f(\theta)
    =
    \frac{(2+\cos\theta)(1-\cos\theta)^2}{4},
    \label{eq:heterogeneous_nucleation_barrier}
\end{equation}
where $\theta$ is the contact angle measured through the vapor phase and
$f(\theta)$ is the geometric factor describing the reduction of the nucleation
barrier due to the solid surface. For $0 \leq \theta \leq \pi$,
$0 \leq f(\theta) \leq 1$, with smaller values of $f(\theta)$ corresponding to
more favorable heterogeneous nucleation sites and a lower free-energy barrier.
Surface roughness, microscopic crevices or cavities, and local thermal or
electrical effects can further modify the effective nucleation barrier and are
not fully represented by this idealized flat-surface contact-angle model~\cite{Fletcher1958,Blander1975}.

To account phenomenologically for the effects of local heating and electric-field-induced interfacial stress, the effective driving pressure for
vapor formation may be written as
\begin{equation}
    \Delta P_{\mathrm{drive}}
    =
    P_v(T_{\mathrm{loc}}) - P
    + \Delta P_{\mathrm{elec}}(E)
    + \Delta P_{\mathrm{other}},
    \qquad
    T_{\mathrm{loc}} = T + \Delta T_J(E),
    \label{eq:effective_driving_pressure}
\end{equation}
where $P_v(T_{\mathrm{loc}})$ is the equilibrium vapor pressure evaluated at the local interfacial temperature $T_{\mathrm{loc}}$, $P$ is the bulk liquid pressure, $\Delta T_J(E)$ represents the local temperature rise associated with field-emission or leakage-current heating, $\Delta P_{\mathrm{elec}}(E)$ denotes an effective contribution from electric-field-induced interfacial or normal stresses, and $\Delta P_{\mathrm{other}}$ represents additional apparatus- or surface-dependent contributions. The corresponding heterogeneous nucleation rate is
\begin{equation}
    \Gamma
    =
    \Gamma_0
    \exp\!\left(
        -\frac{\Delta G_{\mathrm{het}}}
        {k_B T_{\mathrm{loc}}}
    \right),
    \label{eq:nucleation-rate}
\end{equation}
where $\Gamma$ is the nucleation rate for a representative heterogeneous site, $\Gamma_0$ is the corresponding kinetic prefactor, and $k_B$ is the Boltzmann constant~\cite{Turnbull1949}.
Equations~(\ref{eq:homogeneous-barrier})--(\ref{eq:nucleation-rate}) illustrate why modest pressurization can strongly affect the LHe breakdown distribution. For conditions in which $\Delta P_{\mathrm{drive}}>0$, and for fixed local temperature and field-dependent contributions, increasing $P$ reduces $\Delta P_{\mathrm{drive}}$, raises the nucleation barrier, and can strongly
suppress the bubble-nucleation rate. Equation~(\ref{eq:homogeneous-barrier}) is intended for the regime in which
$\Delta P_{\mathrm{drive}}$ remains positive.

Therefore, a plausible interpretation of the observed bimodality is that it arises from a coupled process involving heterogeneous nucleation at favorable electrode-surface sites, field-emission-assisted local heating, and vapor growth. The distinction between the lower- and higher-field populations does not necessarily correspond to homogeneous versus heterogeneous nucleation. Rather, both populations may involve heterogeneous nucleation, with the lower-field component associated with more favorable surface sites or local thermodynamic conditions and the higher-field component associated with sites for which formation or sustained growth of a vapor precursor requires a larger applied field. Once such a vapor cavity forms, field emission and local Joule heating can promote its growth, while the cavity itself provides a low-density path for
charge multiplication. This picture is consistent with the two principal observations: at 602~Torr, the two empirical components exhibit similar field-dependent hazard behavior, whereas pressurization shifts the overall
LHe breakdown distribution to higher field and renders the bimodality less pronounced.

This picture is also consistent with the LHe measurements of Phan~\textit{et al.} at
1.7~K and pressures of 10 and 612~Torr~\cite{Phan2021}. Because both measurements were performed at the same temperature, pressure is the principal thermodynamic distinction between them: the much lower bulk
liquid pressure at 10~Torr provides less external opposition to vapor-cavity growth, whereas pressurization to 612~Torr suppresses it. The effective thermal transport of LHe in the superfluid state is orders of magnitude greater than in the normal-fluid state. This is relevant when comparing these 1.7~K measurements with the present 3.99~K data, but it does not distinguish the two pressure settings
within Ref.~\cite{Phan2021}.

Within this framework, the broad range of LHe breakdown strengths reported in the literature may be interpreted using a common physical picture. Measurements performed near the saturated-vapor-pressure curve may sample different mixtures of the lower- and higher-field breakdown modes, with the relative weight of each contribution depending on electrode topography, surface contamination, thermal gradients, cooldown history, and measurement procedure. Pressurization reduces the susceptibility of the liquid to bubble formation and can therefore suppress the lower-field contribution, providing a more stable reference for comparison between LHe and the other cryogenic liquids studied here. For HV design, the implication is that the mean breakdown field alone is insufficient, because the lower tail of the distribution, especially under conditions conducive to vapor formation, can strongly affect operational reliability.


\section{Summary}\label{sec:summary}

We measured DC electrical breakdown distributions in LN$_2$, LHe, and LAr using a common cryostat, electrode assembly, and voltage-ramp procedure. Use of a common apparatus minimizes several systematic differences that ordinarily complicate cross-liquid studies.

Three liquid-specific behaviors were observed. In LAr, the breakdown voltage increased over the initial sequence of discharges before approaching an approximately stationary plateau. This turn-on behavior is consistent with evolution of the local impurity environment or electrode surface state, although no in situ measurement of impurity concentration at the electrode gap was available. In LN$_2$, the mean breakdown field measured at 760~Torr in the second
cooldown differed from that in the first by approximately 14\% (symmetric fractional difference), indicating sensitivity to electrode-surface evolution even when the geometry was unchanged. In LHe, the saturated-vapor-pressure data exhibited a pronounced lower-field component, whereas the pressurized data shifted to higher field and became less distinctly bimodal. This pressure dependence is consistent with a bubble-mediated contribution to breakdown near saturated vapor pressure.

At approximately 760 Torr, the breakdown fields measured in LN$_2$ and pressurized LHe were comparable within the observed surface-state variation and field-scale uncertainty, whereas LAr broke down at substantially lower fields. This trend is consistent with a phenomenological picture in which lower mobility of the dominant negative charge carrier is associated with greater dielectric strength.

A limitation of the common-electrode approach is that, while it suppresses geometry- and surface-finish differences between liquids, it also introduces run-order dependence through cumulative electrode conditioning, discharge
history, and possible impurity redistribution. The present results show that comparisons based solely on mean breakdown voltage can obscure important distributional structure. In the context of cryogenic HV insulation design, the lower tail of the breakdown-probability distribution and the thermodynamic stability of the liquid phase against bubble nucleation are critical design parameters that
may be as important as, or more important than, the mean breakdown field in determining operational reliability.


\section*{Data availability}

The data that support the findings of this study are available from the corresponding author upon reasonable request, subject to applicable institutional and regulatory requirements.


\acknowledgments

This work was supported by the United States Department of Energy,
Office of Science, Office of Nuclear Physics through Los Alamos National Laboratory under Contract Number 89233218CNA000001, proposal 2023LANLEED3.  The authors are grateful to the Coherent CAPTAIN-Mills experiment for providing the liquid argon utilized in this study.


\section*{Generative AI Statement}
The authors acknowledge the use of generative artificial intelligence tools, specifically GPT 5.4 and 5.6, Gemini 3.1 Flash, and Claude Fable 5, for assistance with literature research and language editing, including improvement of prose flow, grammatical correctness, and formal academic tone. All AI-assisted output was reviewed and verified by the authors and revised where necessary to ensure accuracy, completeness, and consistency with the scientific content of the manuscript. The authors accept responsibility for the integrity and accuracy of the final manuscript.



 \bibliographystyle{JHEP}
 \bibliography{biblio.bib}

\end{document}